\documentclass[varenna,preprint]{cimento}
\usepackage{amsmath}
\usepackage{amssymb}
\usepackage{graphicx}
\usepackage{url}
\usepackage{chngcntr}

\newcommand{\kB}{k_{\text{B}}}
\newcommand{\V}{v}
\newcommand{\PvN}{P_\V(N)}
\newcommand{\PvZero}{P_\V(0)}
\newcommand{\rhocmc}{\rho_{\text{cmc}}}
\renewcommand{\vec}[1]{{\mathbf{#1}}}
\newcommand{\vr}{{\vec r}}

\newcommand{\dee}{\text{d}}

\newcommand{\C}{\mathrm{C}}
\renewcommand{\H}{\mathrm{H}}
\renewcommand{\O}{\mathrm{O}}
\newcommand{\D}{\mathrm{D}}

\newcommand{\onlinecite}{\cite}

\title{Lectures on Molecular- and Nano-scale Fluctuations in Water}
\shorttitle{Fluctuations in water}
\author{David Chandler, Patrick Varilly}
\institute{Department of Chemistry, University of California, Berkeley, Berkeley, CA 94720, USA}

\shortauthor{D. Chandler \atque P. Varilly}

\begin{document}

\maketitle

\begin{abstract}
This paper is the written form of three lectures delivered by one of us (DC) at the International School of Physics ``Enrico Fermi'', Course CLXXVI, ``Complex Materials in Physics and Biology'', held in Varenna, Italy in July 2010. It describes the physical properties of water from a molecular perspective and how these properties are reflected in the behaviors of water as a solvent. Theory of hydrophobicity and solvation of ions are topics included in the discussion.

[Citation: D. Chandler and P. Varilly, in \textsl{Proceedings of the International School of Physics ``Enrico Fermi''}, Vol~\textbf{176}, edited by F.~Mallamace and H.~E.~Stanley (IOS, Amsterdam; SIF, Bologna, 2012), pages 75--111.  \url{http://dx.doi.org/10.3254/978-1-61499-071-0-75}]
\end{abstract}

\section{Lecture One: Tetrahedral Condensed Matter}

Water acts in many ways, from dissolving salt to creating the medium for all life on Earth.  It is so important and multifaceted that whole books can be and are written about it.  See, for example, David Eisenberg's and Walter Kauzmann's timeless and recently re-issued monograph on the physical properties of water~\cite{EisenbergKauzmann2005}. Our charge in these three lectures is to describe something about this material from a molecular perspective.  In limiting scope, we focus on thermal fluctuations and their consequences on solvation and self-assembly.  The presentation is like that of a textbook chapter, not a comprehensive review.  We make use of statistical mechanics at the level it is treated in Ref.~\cite{Chandler1987}.  While our focus is on water, what we say applies to much of liquid matter, where good background is found in Ref.~\onlinecite{BarratHansen2003}. 

The perspective we adopt is influenced by the results of computer simulations because this approach provides unambiguous microscopic detail, albeit for idealized models.  Combined with experiments and theory, simulation is central to all modern understanding of condensed matter.  Water is a most important example. Computer simulations of water were pioneered in the 1970s by Aneesur Rahman and Frank Stillinger.  Many subsequent advances have validated their general approach and enhanced our understanding of water.  Reviews of that early work in Refs.~\cite{Stillinger1976} and~\cite{Stillinger1980} remain informative to this day.

Our first lecture covers properties of pure water, particularly distribution functions related to local arrangements of water molecules.  The second lecture is about free energies of solvation and how these free energies are related to the statistics of spontaneous molecular fluctuations in water.  The third and last lecture builds from that stage to treat forces of self assembly, especially hydrophobic forces, which act on molecular and supramolecular scales.

We begin at the smallest length scales, those of one water molecule.

\subsection{Molecular Structure}
\label{sec:MolStruct}

\begin{figure}
\begin{tabular}{cc}
\includegraphics{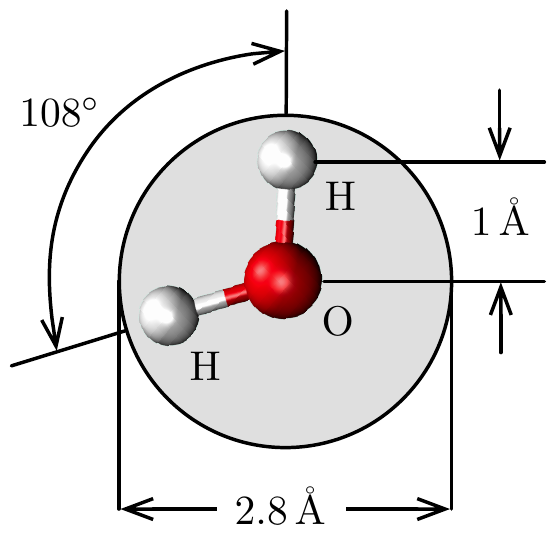}&
\includegraphics{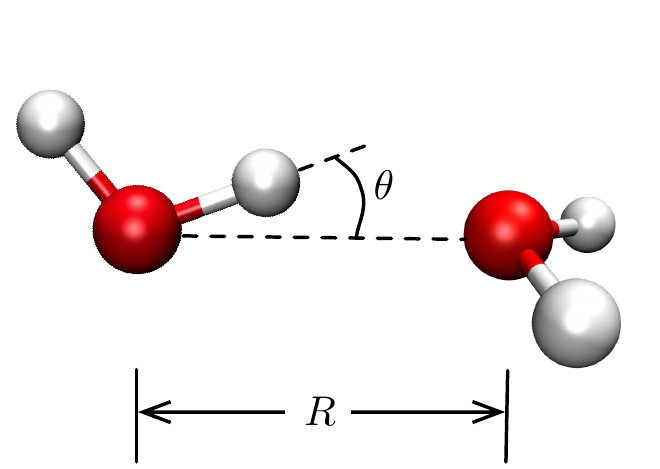}
\end{tabular}
\caption{\label{fig:WaterStructure}Geometry of a water molecule (left) and a typical hydrogen-bond (right).}
\end{figure}

Figure~\ref{fig:WaterStructure} illustrates the water molecule and its most significant interaction---the hydrogen bond.  Though the molecule is quite polar, its electron density is dominated by the electrons of the oxygen atom.  As such, the space-filling volume of a water molecule is approximately spherical with van der Waals radius of~$1.4\,$\AA, like that of its isoelectronic partner, neon.  Because this volume is roughly spherical, it is often convenient to identify the position of a water molecule with the position of its oxygen nucleus.  The $\O\H$ chemical bond is about~$1\,$\AA\ long and the $\H\O\H$ angle is about $108^\circ$.

The hydrogen bond interaction is largely electrostatic in origin, and it is strongest when the intermolecular separation, $R$ in Fig.~\ref{fig:WaterStructure}, is about $3.0 \pm 0.2\,$\AA, and the angle, $\theta$, is $\theta \lesssim 20^\circ$.  This linear hydrogen-bond has a maximum adhesive strength of about $6\,$kcal/mol, which coincides with~$10\,\kB T$ at room temperature (i.e., $25^\circ\,$C).  Such a large attraction between molecules is unusual.  But the strength is only this high over a limited range of relative positions and orientations, so that most of the hydrogen-bond strength is lost for $\theta \gtrsim 30^\circ$ and $R \gtrsim 3.5\,$\AA, and the interaction decays as~$1/R^3$ to negligible values for separations that are a few Angstroms larger.  

Liquids in general are characterized by nearly-balanced competition between energy and entropy.  When energy dominates, the material is solid, and when entropy dominates, the material is gaseous.  Liquid water exhibits this balance and is thus ordinary in this and many other respects.  Its values for density, viscosity and diffusion constant are comparable to those of other liquids at ambient conditions (i.e., $25^\circ\,$C and $1\,$atm pressure). But the energy-entropy competition in water is remarkable because its large adhesive energy acts over a limited range of configurations.  It makes water fragile in the sense that its properties have unusually strong dependence with respect to temperature and pressure, as discussed below.

\begin{figure}
\begin{tabular}{cc}
\includegraphics[width=60mm]{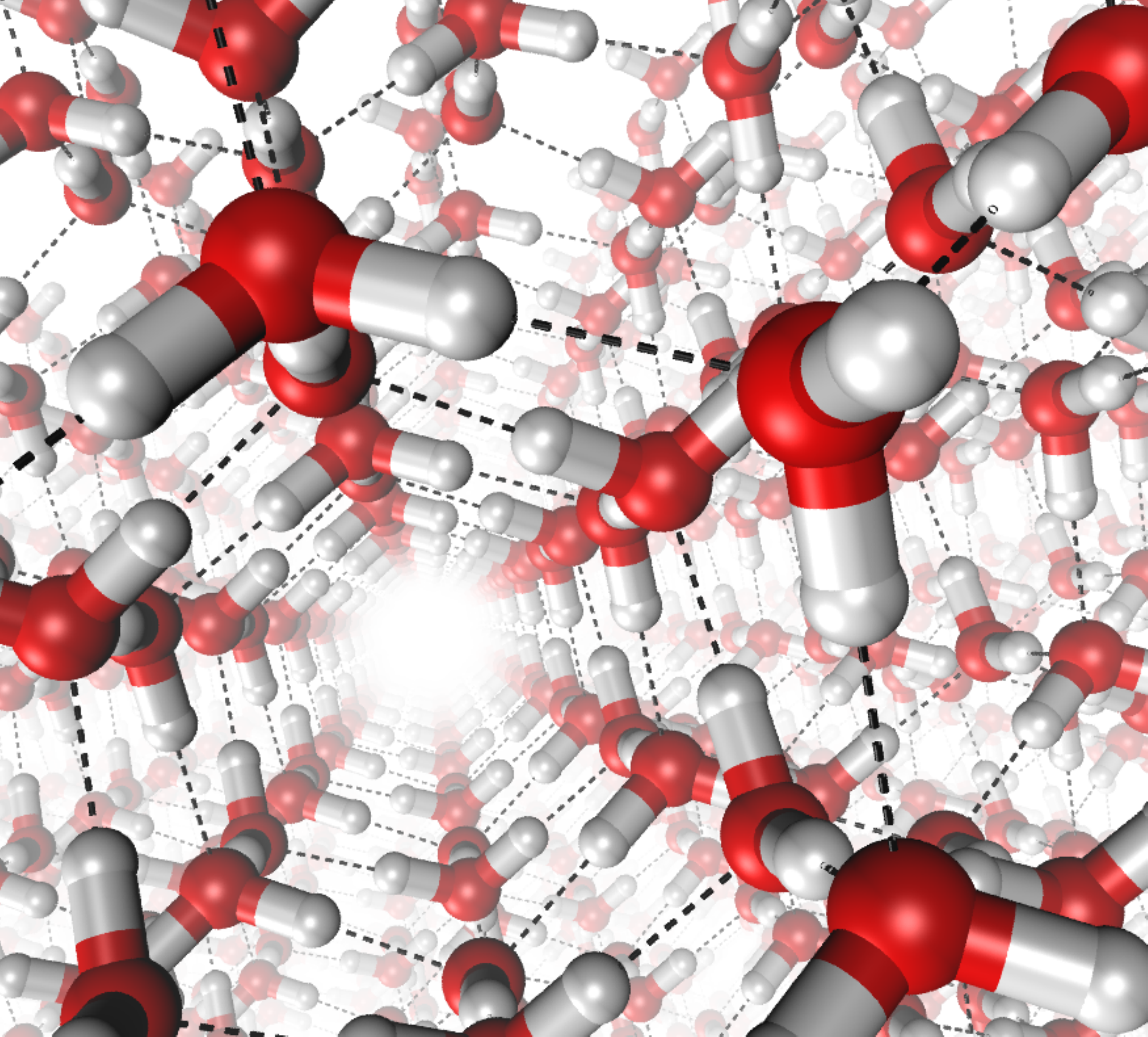}&
\includegraphics[width=60mm]{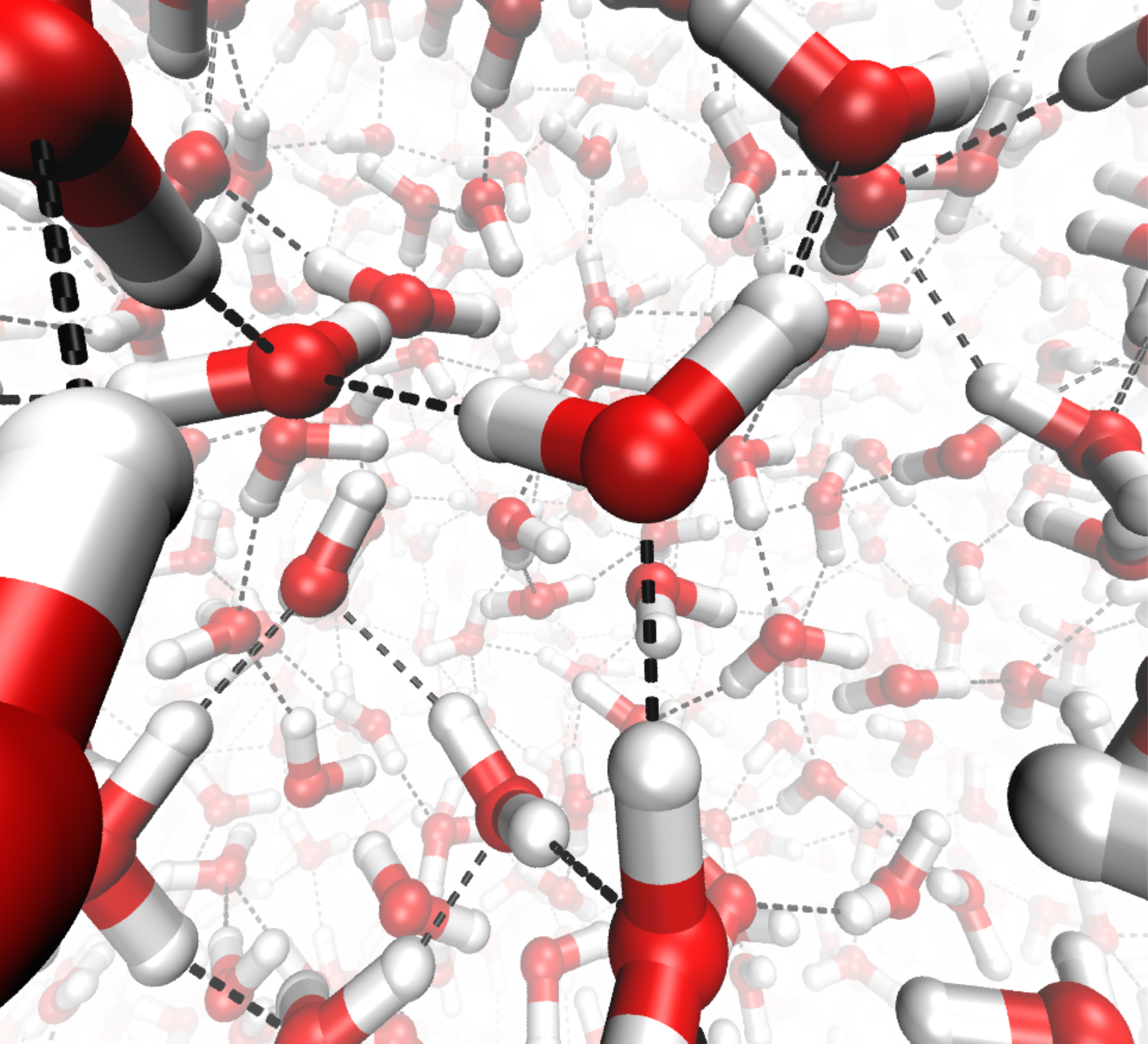}\\
(a) Ice&(b) Liquid Water
\end{tabular}
\caption{\label{fig:HBondNetwork}Hydrogen-bonding networks of condensed water, where dashed lines depict hydrogen bonds.  The picture on the left is a rendering of the molecular structure of ice I.  The picture on the right is a configuration generated through a classical trajectory (i.e., molecular dynamics computer simulations) in which forces are computed according to the SPC/E model of intermolecular potentials~\cite{BerendsenGrigeraStraatsma1987}.  Measures of structure, dynamics and thermodynamics obtained with this and similar models are in reasonable harmony with those obtained from experiments on real water.}
\end{figure}

Frozen water is ice, depicted in Fig.~\ref{fig:HBondNetwork}(a), with the individual molecules arranged in nearly perfect tetrahedral order.  The resulting hydrogen-bonding network is nearly defect-free, with each water hydrogen-bonded to four other waters.  Liquid water, on the other hand, is bound together by a disordered network of hydrogen-bonds, depicted in Fig.~\ref{fig:HBondNetwork}(b).  Hydrogen bonds are constantly breaking and reforming, so that hydrogen-bonded partners switch allegiances regularly.  

On average, a water molecule in the liquid will have fewer than the energetically optimum four hydrogen-bonding partners.  But the number of hydrogen bonds fluctuates, and for a tagged water molecule at any one point in time the number typically ranges from two to five.  On rare occasions, there are even as few as one or none.  The precise numbers and their probabilities depend upon the definition of a hydrogen bond, and conventions for that definition are somewhat arbitrary.   For any reasonable definition, however, a basin in the potential energy landscape of liquid water coincides with a specific hydrogen-bonding network, whereas transition states or saddles separating basins in that landscape coincide with breaking hydrogen bonds in the network, and the probability distribution for the number of hydrogen bonds that a given molecule experiences is unimodal.

Without theory and simulation, experimental probes of water structure and dynamics are often difficult to interpret.  This is especially true for spectroscopy.  These experiments occasionally motivate pictures of water structure different than those described above.  One example is the two-state model of water, where the liquid is imagined to be an ideal mixture of bonded and non-bonded molecules.  Unions of experiment with theory and simulation have dispensed with that idea, as illustrated, for example, in Refs.~\cite{FeckoEtAl2003}, \cite{EavesEtAl2005} and \cite{Geissler2005}. 

One experimentally-accessible measure of liquid structure is a pair distribution function.  For water, there are three: $g_{\O\O}(r)$, $g_{\O\H}(r)$~and~$g_{\H\H}(r)$, which measure the probability that two different atoms, oxygen~or~hydrogen, are a distance~$r$ apart.  More precisely,
\begin{equation}
\label{g_of_r}
\rho g_{\O\O}(r) = V \Bigl\langle \sum_{j>1} \delta(\vr_1^{(\O)}) \delta(\vr_j^{(\O)} - \vr) \Bigr\rangle,
\end{equation}
where $\rho$ is the number density of water, $V$ is the volume of the system, $\vr_i^{(\O)}$ denotes the position of the oxygen atom of molecule~$i$, the angular brackets denote a thermal average, and the $\delta$-functions are Dirac's functions, which have unit volume and are non-zero only at points where their arguments are zero.  Analogous expressions apply for $g_{\O\H}(r)$~and~$g_{\H\H}(r)$.  In an isotropic system, they depend on only the magnitude of $\vr$, $r = |\vr|$.

\begin{figure}\begin{center}
\begin{tabular}{cc}
\includegraphics{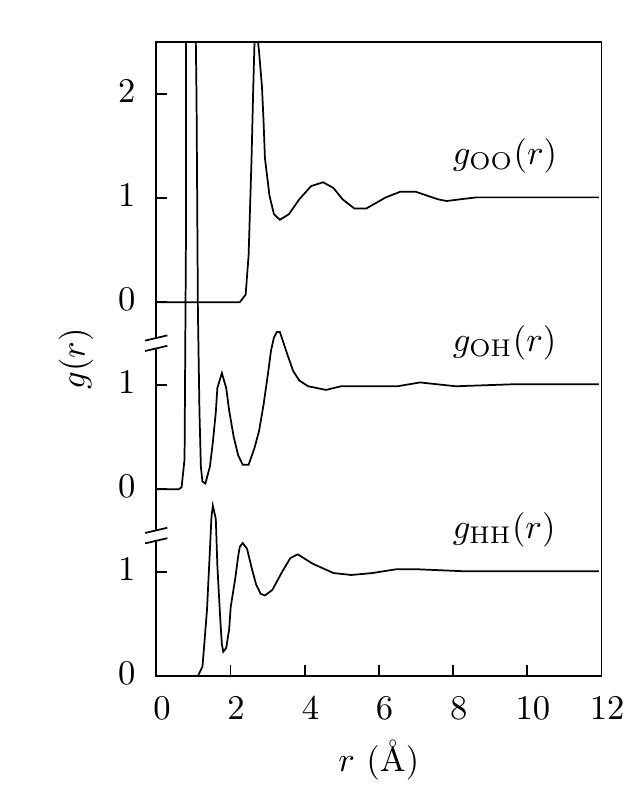}&
\includegraphics{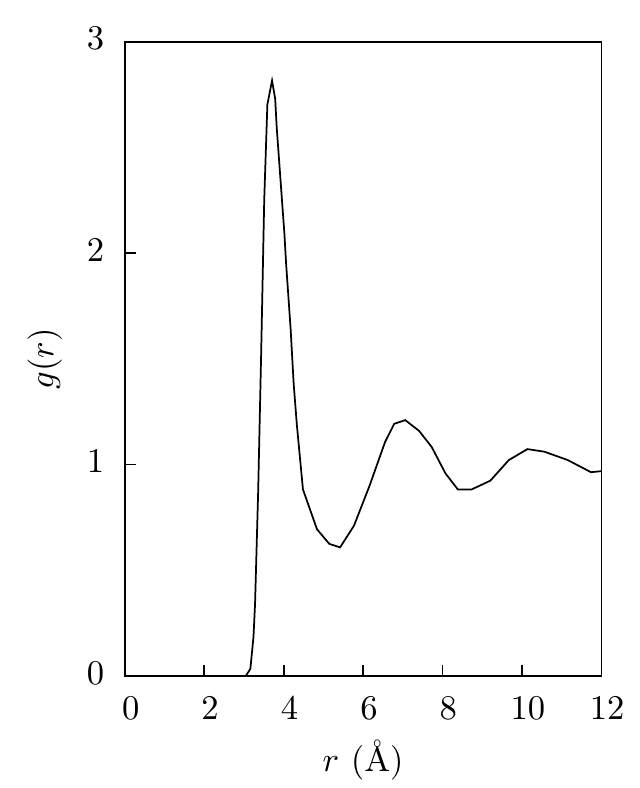}\\
\includegraphics{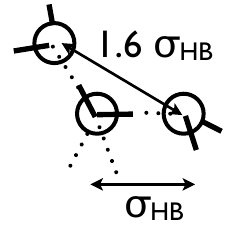}&
\includegraphics{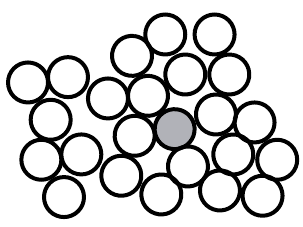}
\end{tabular}
\end{center}\caption{\label{fig:neutronGAndLJG}Left: Pair distribution functions of water at ambient conditions, as obtained from neutron scattering experiments (adapted from Ref.~\cite{Soper2001}).  The shortest distance peaks for the OH and HH functions refer to  intramolecular separations between atoms.  Neutron scattering detect these peaks along with intermolecular features.  The intramolecular peaks are shown here for purposes of comparing with remaining intermolecular features.  Right: Radial distribution function of a model of Argon at temperature $T=94.4^\circ\,$K and density $\rho=1.374\,$g/cc (adapted from Ref.~\cite{Rahman1964}).  The schematic pictures at the bottom depict the pertinent geometries for water (left) and for argon (right) from which the main features in the exhibited distribution functions can be rationalized.}
\end{figure}

Neutron and x-ray scattering experiments probe different weighted combinations of these functions.  By adjusting isotopic concentrations and thereby altering weights, each distribution function may be estimated.  Figure~\ref{fig:neutronGAndLJG}(a) shows the three pair distribution functions as deduced from neutron scattering measurements.  These functions are fully consistent with and supportive of a picture in which liquid water is organized with a preference for local tetrahedral ordering.

For comparison, Fig.~\ref{fig:neutronGAndLJG}(b) shows the radial distribution function of argon.  Argon has no strong and directional adhesive forces, and it is therefore called a non-associated liquid.  Most liquids are non-associated.  Structure in such liquids is dominated by packing, and since argon atoms are spherical, liquid argon structure is essentially that of a dense fluid of hard spheres.  An integral over the first peak in $g(r)$ indicates that an atom in liquid argon has on average about 12 nearest neighbors, while the corresponding integral over $g_{\O\O}(r)$ indicates that a water molecule in liquid water has on average about 4 nearest neighbors.  Further, while the second peak for the $g(r)$ of liquid argon is at about twice the nearest-neighbor separation, the second peak of $g_{\O\O}(r)$ is located at around $1.6$~times the nearest-neighbor separation, corresponding to water molecules that share a common hydrogen-bonding partner with local tetrahedral order.

\begin{figure}\begin{center}
\includegraphics{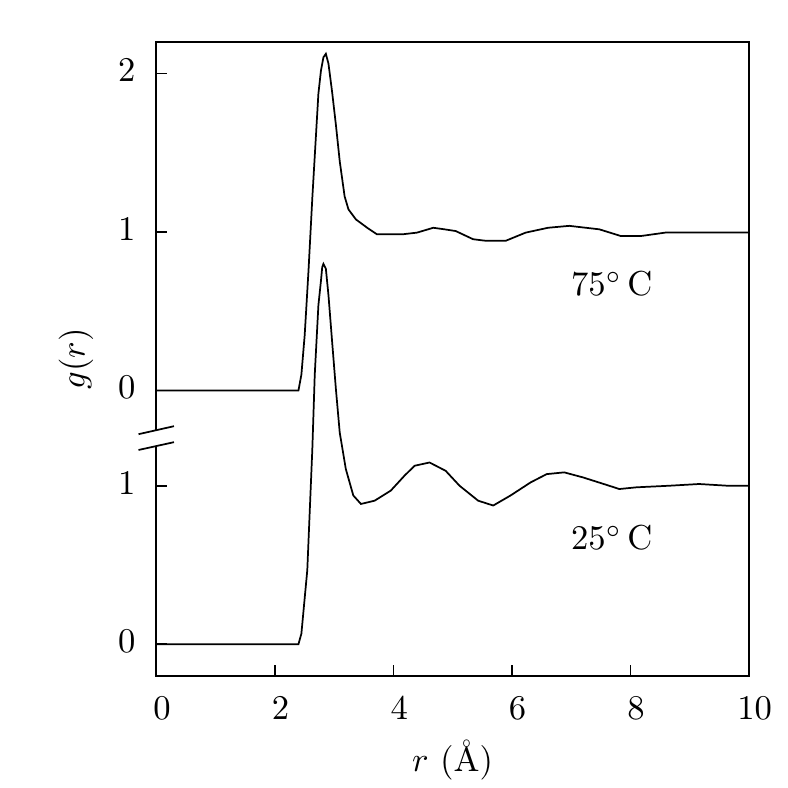}
\end{center}\caption{\label{fig:XRaysGofR}Pair distribution functions of water at $25^\circ\,$C~and~$75^\circ\,$C, determined from X-Ray scattering measurements.  Adapted from Ref.~\onlinecite{NartenLevy1971} (data smoothed for clarity).}
\end{figure}

Because non-associated liquids are dominated by packing forces, as detailed in Weeks-Chandler-Andersen theory~\cite{ChandlerWeeksAndersen1983}, the structures of these fluids are nearly athermal, responding to changes in temperature only insofar as these cause changes in density.  The structure of water, on the other hand, is very sensitive to temperature.  Figure~\ref{fig:XRaysGofR} shows the function~$g_{\O\O}(r)$ for water, as determined from X-ray scattering experiments at~$25^\circ\,$C~and~$75^\circ\,$C. The most striking feature is that most of the structure beyond the first peak disappears above around $50^\circ\,$C, even though the density of the liquid decreases by less than~$1$\%.  This notable change can be rationalized in terms of the competition between entropy and enthalpy.  While hydrogen-bonds are energetically very favorable, the open structure they impose is entropically unfavorable.  At ambient conditions, enthalpy narrowly dominates, while at higher temperatures, entropy plays the larger role.  In warm water, above $50^\circ\,$C, hydrogen bonds are often broken and no longer dictate local tetrahedral ordering of molecules.  This sensitivity to temperature implies that water has a large heat capacity at ambient conditions.  Indeed, per water molecule, the heat capacity at constant volume, $C_{\text{\tiny V}}$, is around $10\,\kB$.  For simple non-associated liquids, the corresponding heat capacity is an order of magnitude smaller.

\begin{figure}
\begin{tabular}{cc}
\includegraphics{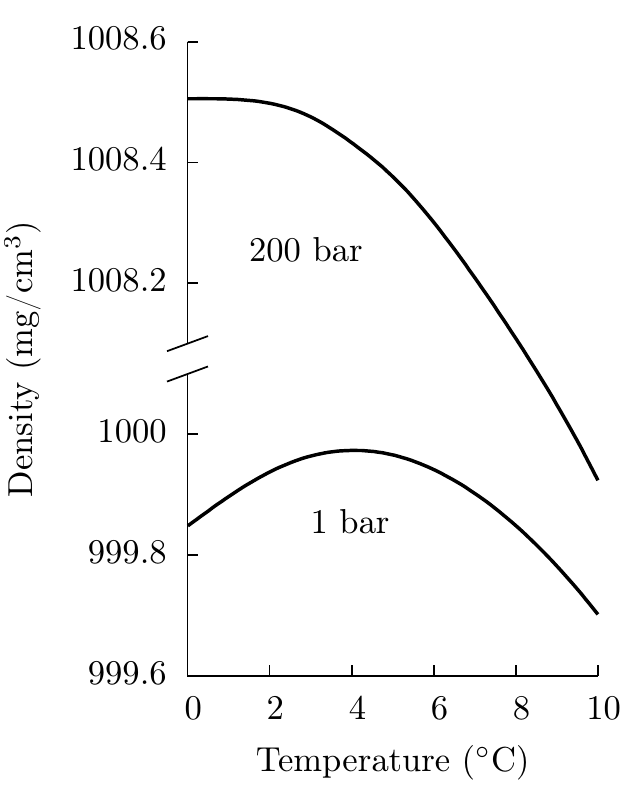}&
\includegraphics{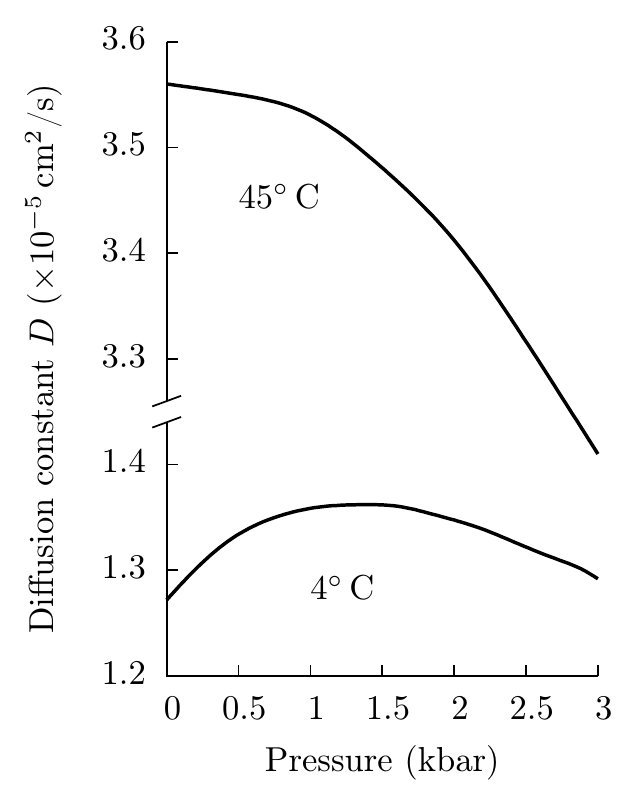}\\
(a)&(b)
\end{tabular}
\caption{\label{fig:DensityAndDiffusionMaximum}(a)~Density of water at $1$~bar and $200$~bar.  Data from~\onlinecite{NISTWater}.  (b)~Self-diffusion constant of water at $T=4^\circ\,$C~and~$T=45^\circ\,$C.  Adapted from Ref.~\onlinecite{HarrisWoolf1980}.}
\end{figure}

The competition between hydrogen-bond enthalpy and packing entropy is manifest in many other properties of water around ambient conditions, though the specific crossover temperatures and pressures vary slightly with different observables.  For example, at atmospheric pressure, water attains a maximum density at~$4^\circ$C, where its structure is most open (Fig.~\ref{fig:DensityAndDiffusionMaximum}(a)).  Other examples are discussed below.

\subsection{Diffusion}
\label{sec:diffusion}

Fluctuations in the structure of liquid water allow molecules to diffuse.  The temperature and pressure dependence of the self-diffusion constant of water reflects the degree to which fluctuations change with external conditions.  For temperatures below about $310\,$K, there exists a pressure at which the diffusion constant~$D$ of water is maximal, as shown in Fig.~\ref{fig:DensityAndDiffusionMaximum}(b).  As with the density maximum, this maximum is a manifestation of the nearly-balanced competition between enthalpy and entropy, about which we will have more to say.  First, however, note that the diffusion constant of water under most common conditions is, to within an order magnitude, equal to $10^{-5}\,$cm$^2$/s, a value that is typical of dense liquids at ambient conditions.  This constant measures the rate at which the mean-squared displacement of a particle increases with time,
\begin{equation}
\langle | \vr_1^{(\O)}(t) - \vr_1^{(\O)}(0) |^2 \rangle \sim 6 D t,\quad \text{for }t\,\,\, \text{large}.
\end{equation}
The value of $D$ for water is such that after about~$1\,$ps, on average a water oxygen travels about~$1\,$\AA.  By taking the time-derivative of both sides of the above equation, we obtain the relationship connecting the self-diffusion constant to the velocity auto-correlation function,
\begin{equation}
\beta m D = \langle {\vec v}^2 \rangle^{-1} \int_0^\infty dt\, \langle \vec{v}(0)\cdot\vec{v}(t) \rangle = \tau_v \approx 10^{-12}\,\text{s}.
\end{equation}
Here, $\beta^{-1} = \kB T$, $m$ is the mass of an oxygen atom, and $\vec{v}$ is shorthand for $\vec{v}_1^{(\O)}$, the velocity of the oxygen atom of water molecule~$1$.
The right-hand side of this equation is the time-scale~$\tau_v$ beyond which $\langle | \vr_1^{(\O)}(t) - \vr_1^{(\O)}(0) |^2 \rangle$ goes as $6Dt$, and this time for water, of the order of 1 ps, is also typical of most dense liquids at ambient conditions.

The diffusion constant is a measure of average molecular motion.  The actual motions of liquid water molecules are intermittent, and at any single 1 ps time frame, a tagged water may or may not move a distance of the order of 1 \AA.  For a water molecule to diffuse, it must break an existing hydrogen bond and form a new bond.  Conversely, after a hydrogen bond is broken, the possibility of the previously-paired water molecules to drift apart is much larger than when they are bonded.  Diffusion and hydrogen-bond dynamics are thus coupled.  At least in part because of this coupling, hydrogen-bond lifetimes have a non-trivial distribution~$P_{\text{HB}}(\tau)$, whose mean corresponds to the typical bond lifetime,
\begin{equation}
\tau_{\text{HB}} = \int \tau \,P_{\text{HB}}(\tau) \, d\tau \approx 10^{-12}\,\text{s}.
\end{equation}
Direct experimental measures of $P_{\text{HB}}(\tau)$ are not available.  Its general features can be inferred from simulation.  The typical lifetime~$\tau_{\text{HB}}$ is comparable to the diffusion timescale~$\tau_v$.  

The maximum in the diffusion constant reflects the effect of pressure on the entropy-enthalpy balance of the hydrogen-bonding network of water.  As pressure increases from $1\,$bar to about $1\,$kbar, hydrogen bonds are destabilized, so it is easier for water molecules to break and reform hydrogen bonds and thus, diffuse.  For moderate pressures, this effect is larger than the counteracting slowdown in diffusion expected from the higher densities induced by higher pressures.  There is a crossover at around $1\,$kbar, above which the diffusion constant, as in a simple liquid, decreases with increasing pressure.  At high temperatures, as evidenced in Fig.~\ref{fig:XRaysGofR}, the hydrogen-bonding network is already substantially perturbed, and only the simple-liquid dependence of diffusivity on pressure is observed.

\subsection{Chemistry in water}
Water plays important roles as the medium or solvent for chemical processes.  It also plays important roles as a reactant.  Chemical properties of liquid water in those cases mostly pertain to the presence of disassociated water molecules, the hydroxide and hydronium ions, $\H\O^-$ and $\H_3\O^+$, respectively.  They are in dynamic equilibrium with intact water molecules:   $2\,\H_2\O   \rightleftharpoons \H\O^- + \H_3\O^+$. While the concentrations of these ions are extremely low (about $10^{-8}$~times smaller than that of intact water molecules), the presence of these ions is crucial to acid-base chemistry, biological pumps and motors, and, possibly, transportation fuels of the future.  

A proton in water, sometimes denoted by~$[\H^+]_\mathrm{aq}$, is usually bonded to or strongly associated with one or more water molecules.  Most often it appears in a hydronium ion, $\H_3\O^+$, surrounded by other water molecules.  Such protons diffuse in water faster than do the individual molecules.  They do so by moving along hydrogen-bonding wires, whereby a hydronium ion ($\H_3\O^+ = [\H^+]_{\mathrm{aq}}$) that is hydrogen-bonded to a water molecule ($\H_2\O$) transfers the extra proton along a hydrogen bond.  Recall that in a hydrogen bond, the chemical $\O$-$\H$ bond on one molecule points towards the O of another molecule, $\O\H \cdots \O$.  The H in that arrangement is called the donor hydrogen, and it is a proton of a donor hydrogen that moves from the $\H_3\O^+$ to the adjacent $\H_2\O$.  After this transfer, the initially donating ion becomes a stable water molecule and the initially receiving water molecule becomes an ion.  The new ion can then transfer one of its three protons in a similar fashion to yet another water molecule.  This chain of events, called the Grotthus mechanism, can continue indefinitely provided water molecules are networked together with hydrogen bonds.  Through this mechanism, excess charge of the hydronium ion can travel with minimal diffusion of water molecules. 

While the diffusion of the proton is faster than that of intact water molecules, it is only so by factors that are of order one.  This is because the shuttling of charge along a chain of hydrogen bonds requires forces from molecules surrounding the chain, and these in turn require some changes in the orientations of water molecules---the same sorts of reorganization that can also lead to hydrogen bond breaking and molecular diffusion.  Illustrative computer simulations of this dynamics of water and solvated protons in Refs.~\cite{SchmittVoth1999}, \cite{DayEtAl2002}~and~\cite{MarxEtAl1999} employ models in which species can break and make chemical bonds.  

Auto-ionization of water, $[\H_2\O]_{\mathrm{aq}} \rightarrow [\H^+]_{\mathrm{aq}} +[\O \H^-]_{\mathrm{aq}}$, also involves the Grotthus mechanism, as shown in Fig.~\ref{fig:AutoIonization} below.  But before describing how this ionization occurs, we consider further the sorts of fluctuations that occur in water.  It is fluctuations---in this case, fluctuations of electric fields---that can cause a proton to leave one oxygen and join another.  

\subsection{Density fluctuations}
\label{sec:DensityFlucts}

Microscopic fluctuations in density are related to how water adjusts to the presence of solutes in general, and these fluctuations control the organization of apolar molecules in water in particular.  Remarkably, density fluctuations on small length scales obey Gaussian statistics to good accuracy.  We will exploit this fact in the next two lectures.  

\begin{figure}\begin{center}
\begin{tabular}{cc}
\includegraphics{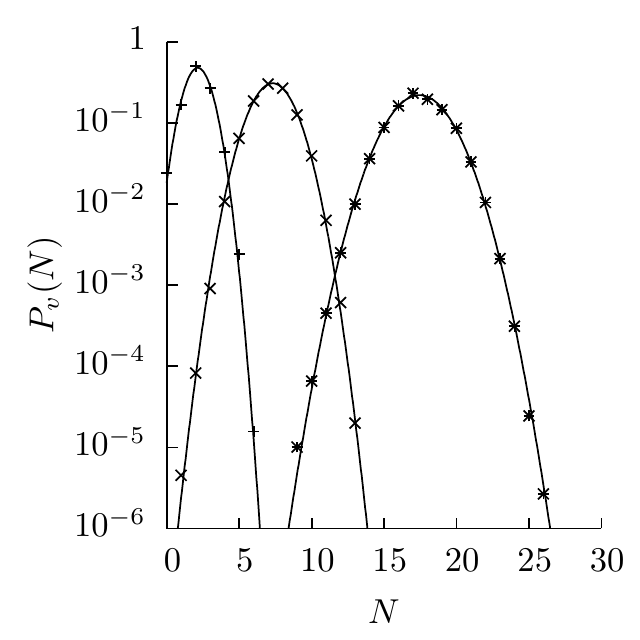}&
\includegraphics{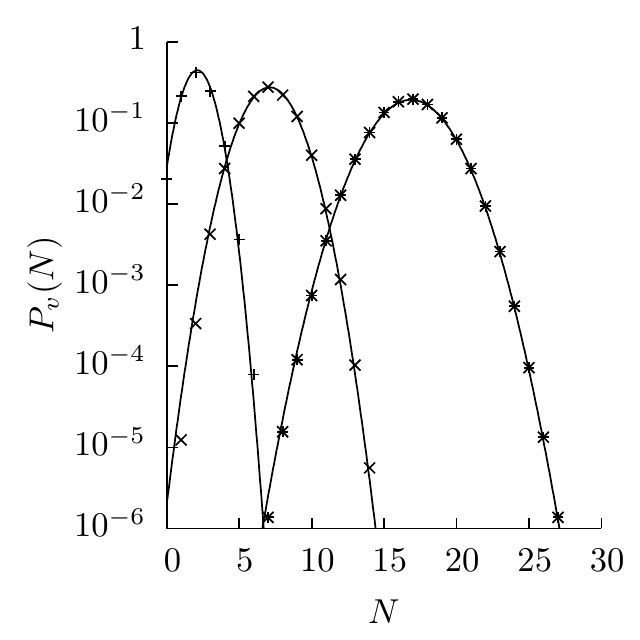}\\
(a) Liquid Water&(b) Hard Sphere Fluid
\end{tabular}
\end{center}
\caption{\label{fig:GaussianDensities} (a) Probability of seeing $N$ waters in a spherical volume $\V$ of radius (left to right) $2.5\,$\AA, $3.75\,$\AA~and~$5.0\,$\AA.  Simulation results at ambient conditions (symbols) are compared with a Gaussian of equal mean and variance (lines).  After Ref.~\onlinecite{HummerEtAl1996}.  (b) Same, for a fluid of hard spheres of diameter~$\sigma$ at density~$\rho=0.5\,\sigma^{-3}$, with a spherical probe volume $\V$ of radius (left to right) $\sigma$, $1.5\sigma$~and~$2\sigma$.  Adapted from Ref.~\onlinecite{CrooksChandler1997}.}
\end{figure}

One way to study density fluctuations is to use computer simulations to calculate the probability $\PvN$ for observing $N$~waters within a probe volume~$\V$.  Figure~\ref{fig:GaussianDensities}(a) shows these probabilities for water in spherical probe volumes, and compares them to Gaussian distributions with the same mean and variance.  Gaussian statistics are not unique to water.  Figure~\ref{fig:GaussianDensities}(b) shows a similar collection of $\PvN$ distributions in a hard-sphere fluid.  Gaussian statistics of this sort are consistently found in numerical simulations of a wide variety of dense homogeneous fluids.  This result is remarkable because it is not obvious why the central limit theorem should apply for molecular-scale probe volumes~$\V$.

For large probe volumes~$\V$, mean-square density fluctuations are related to the isothermal compressibility.  In particular,
\begin{equation}
\label{compressibility}
\frac{\langle(\delta N)^2\rangle_\V}{\langle N \rangle_\V} \to \frac{\partial \rho}{\partial \beta p}\,,\qquad \V\to\infty\,,
\end{equation}
where $\rho$ is the average (bulk) molecular density, $p$ is pressure, and $\delta N = N - \langle N \rangle_\V$ is the fluctuation in the number of molecules in~$\V$, $N$.  Equation~\eqref{compressibility} is a standard result of the grand canonical ensemble; see, for instance, Refs.~\cite{Chandler1987} and~\cite{HansenMcDonald2006}.  The limiting value of Eq.~\ref{compressibility}, $\partial\rho/\partial\beta p$, is a unit-less characterization of the compressibility.  It is 1 for an ideal gas, and it far exceeds 1 and ultimately diverges as a fluid approaches a critical point.  But liquid water at standard conditions is not an ideal gas and it is far from a critical point.  Rather, it is near its triple point.  For typical liquids near triple points, $\partial \rho / \partial \beta p$ is of the order of $10^{-2}$.  For water, its value is about $0.06$, reflecting the not atypical but relatively open and malleable structure of this liquid.  

From Eq.~\ref{compressibility}, one expects that the variance of $P_\V(N)$ will increase as the size of the probe volume $\V = \langle N \rangle_\V / \rho$ increases.  This behavior is indeed observed in the probability functions shown in Fig.~\ref{fig:GaussianDensities}.  But for volumes $\V$ smaller than about 1 nm$^3$, the mean-square size of fluctuations in $N$ is larger than that estimated from Eq.~\ref{compressibility}.  In other words, homogeneous water is generally stiffer on large length scales (more than 1 nm) than on small length scales (less than 1 nm).  In the next lecture, we have more to say about this fact, which pertains to the correlation length of water.

\subsection{Electric field fluctuations}
\label{sec:EFieldFlucts}

Microscopic fluctuations in electric fields play a prominent role in charge transfer reactions.  In such processes, charge initially localized on one site transitions to be localized at a different site.  These transferring charges can be associated with an electron or a proton.  When the transition occurs spontaneously, it is the result of thermal fluctuation.  In particular, these fluctuations can produce electric potentials that make the two localized charge states isoenergetic, facilitating movement of charge from one state to the other.  The activation energy for this process is therefore dominated by the energies of the facilitating fluctuation in electric potentials, which ultimately refers to the free energies and therefore probabilities for fluctuations in arrangements of water molecules.  These ideas form the basis of Rudolph Marcus' celebrated theory of electron transfer.  An extended introduction to that theory is found in Ref.~\onlinecite{Chandler1998}, and in Marcus' Nobel Prize lecture published in Ref.~\onlinecite{Marcus1993}.

\begin{figure}\begin{center}
\begin{tabular}{cc}
\includegraphics[width=6cm]{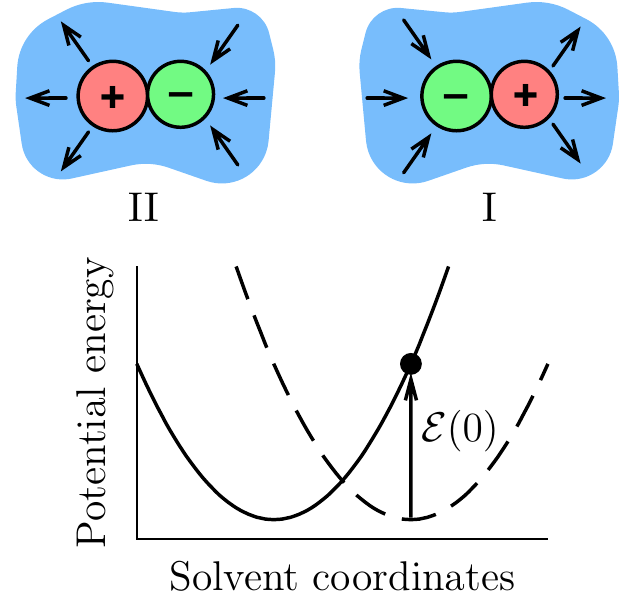}&
\includegraphics[width=7cm]{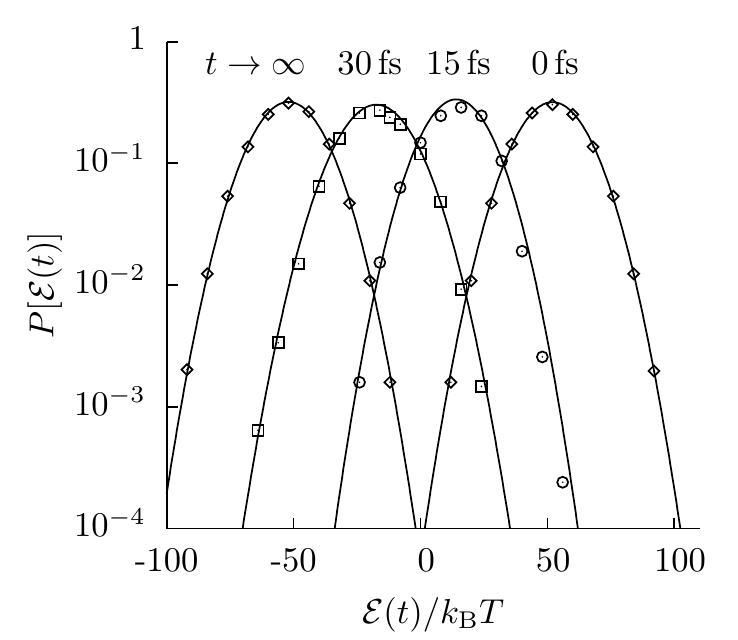}\\
(a)&(b)
\end{tabular}
\end{center}\caption{\label{fig:Marcus}(a) Schematic of a solute in either state I~or state~II and the associated energies of interaction with the surrounding water.  Photoexcitation promotes a transition from I~to~II at time~$t=0$ (following Ref.~\cite{JimenezEtAl1994}).  (b) Probability of observing an energy gap~$\mathcal{E}$ a time~$t$ after photoexcitation (symbols: simulation, lines: Gaussian with equal mean and variance).  Adapted from Ref.~\onlinecite{GeisslerChandler2000}.}
\end{figure}

The basic idea of Marcus theory is shown in Fig.~\ref{fig:Marcus}(a).  The potential energy of two charge states, denoted I~and~II, is shown as a function of the structure of the solvent.  Generically, given a particular solvent configuration, the energies of the two charged states will differ by an amount~$\mathcal{E}$, called the energy gap.  Without an external perturbation, the activation barrier to charge transfer is given by the free energy to change the solvent structure from a typical configuration to one where the two potential energy surfaces intersect.  

In experiments, rather than wait for charge transfer to happen spontaneously (i.e., in the dark), one can inject a photon of energy~$\mathcal{E}(0)$ to produce a transition at time~$t=0$ from electronic state~I to electronic state~II.  In computer simulations, one can take a typical equilibrium configuration of state~I and instantaneously change the charge configuration to coincide with that of state~II.  In both cases, one can subsequently monitor the relaxation of the surrounding solvent as it adapts to the presence of the new charge state.  Figure~\ref{fig:Marcus}(b) illustrates the statistics of the energy gap as a function of time elapsed since the charge transfer reaction.  Here too, Gaussian statistics prevails, in this case for the energy gap, not only at equilibrium, but at all times during the non-equilibrium relaxation after the charge transfer.  

Another important observation illustrated in Fig.~\ref{fig:Marcus}(b) is the speed at which the excess energy added to the system at~$t=0$ is dissipated into the solvent.  In this example, about $60\,\kB T$ is injected into the system.  Within $30\,$fs, two thirds of that energy has been transferred to the solvent.  The fast dynamics illustrated in Fig.~\ref{fig:Marcus} are polarization fluctuations associated with small rotational or rocking motions of water molecules---termed librations.  These motions occur with high frequencies because protons have small mass and thus water molecules have small moments of inertia.  

Because librations are so rapid, nuclear quantum effects can be important.  Any motion of frequency~$\omega$ such that $\hbar\omega / \kB T \gtrsim 1$ will show quantization effects, and at ambient conditions, $\hbar / \kB T \approx 25\,$fs.  Indeed, one mechanism of polarization fluctuations in water is through quantum tunneling. Librations move molecules through classically forbidden regions of configuration space.  If the hydrogens are replaced by deuteriums, tunneling probabilities decrease because tunneling probabilities in general decrease exponentially with increasing particle mass.  This decrease explains some of the substantial isotope effects in electron transfer reactions.  Other observable nuclear quantum effects include a small difference in melting temperatures ($4^\circ$C for $\D_2\O$, $0^\circ$C for $\H_2\O$) and the temperature of maximum density ($12^\circ$C for $\D_2\O$, $4^\circ$C for $\H_2\O$).  Theory for such effects is described in Refs.~\onlinecite{Chandler1991} and \onlinecite{Nitzan2006}.

\subsection{Water auto-ionization}
\label{sec:AutoIonization}
\begin{figure}\begin{center}
\includegraphics[width=130mm]{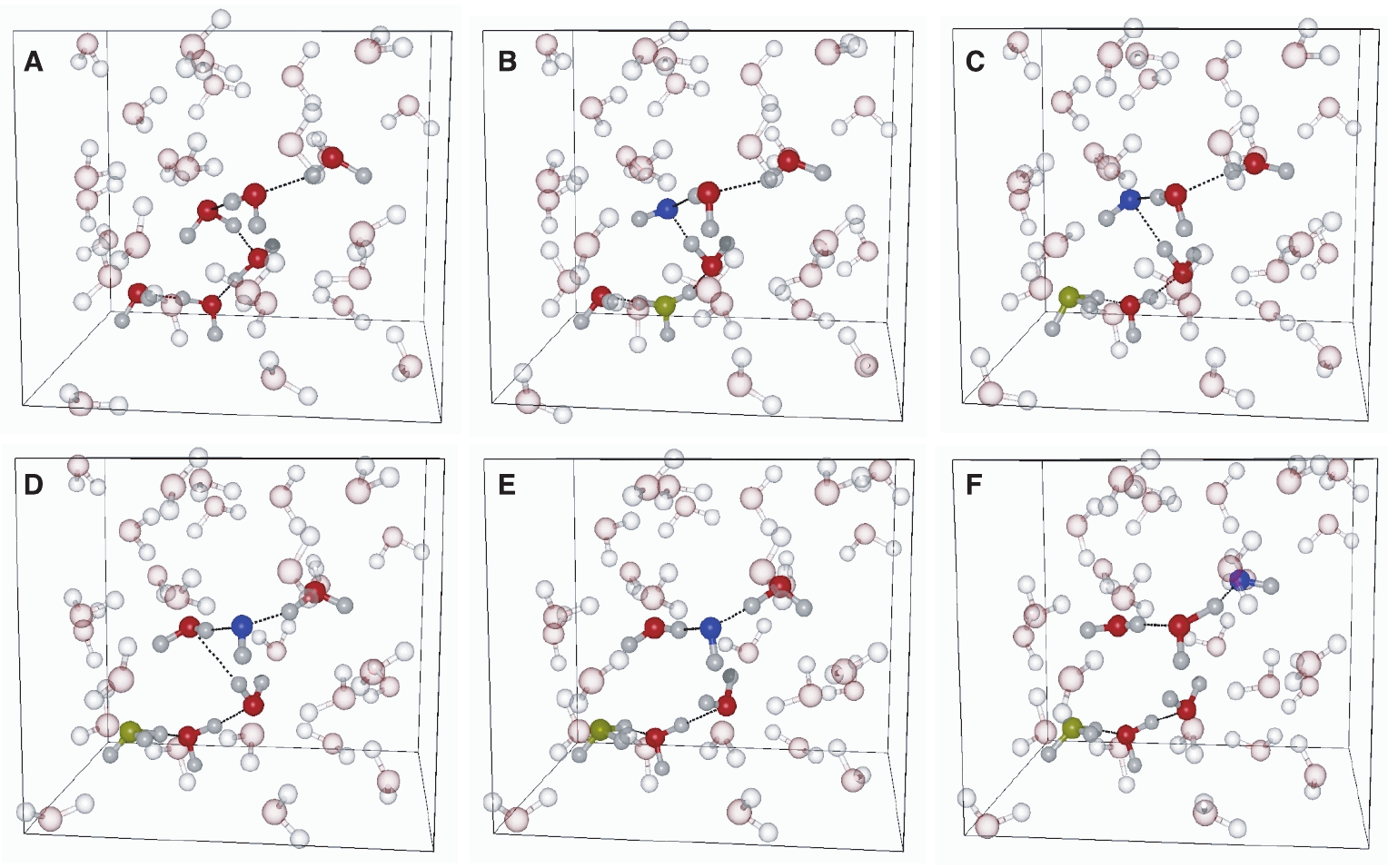}
\end{center}\caption{\label{fig:AutoIonization}Typical $150\,$fs trajectory illustrating an auto-ionization event, from Ref.~\onlinecite{GeisslerEtAl2001}.  Each panel is separated by $30\,$fs.  The transition state for the process is visited in a time frame between those pictured in Panels D and E.  The hydroxide and hydronium ions are shown in blue and yellow, respectively.  The subset of hydrogen bonds that play a specific role in this particular trajectory are shown with dotted lines.}
\end{figure}

To conclude this first lecture, we discuss the role of fluctuations on the mechanism of water auto-ionization.  At equilibrium in water, once in roughly 10 hours, a tagged intact water molecule will dissociate to become a hydroxide ($\O\H^-$) ion, thereby giving up one of its protons to the surrounding liquid.  The process is both rare (occurring only once in hours) and fleeting (completing in only picoseconds after it starts).  Such events can be examined in computer simulations through an importance sampling of trajectory space.  This sampling was used to harvest many independent examples of auto-ionization events in water, and one example is shown in Fig.~\ref{fig:AutoIonization}.  From analyzing the harvested trajectories and characterizing their transition states, it has been found that auto-ionization follows from the coincidence of two fluctuations, which we describe now. 

First, a large electric field fluctuation briefly destabilizes an $\O$-$\H$ chemical bond and thus stabilizes an ion-separated configuration.  Fluctuations large enough to break chemical bonds occur in water, but they are extremely rare.  They are unusual circumstances where electric (mostly dipolar) fields from many, many molecules focus at the site of one O-H bond.  But when this happens, because molecular orientations change quickly, the large destabilizing electric field lasts for no more than a few tens of femtoseconds.  The presence of such large and fleeting electric fields in water was discovered by Graham Fleming and his co-workers~\cite{JimenezEtAl1994}.  For the brief period of time when this large field persists, an O-H chemical bond is destabilized and charge separation can occur along a hydrogen bond wire by the Grotthus mechanism, as seen in Panels B, C and D of Fig.~\ref{fig:AutoIonization}.  

A second fluctuation is required to break the wire of hydrogen bonds along which the charge separation proceeds by the Grotthus mechanism.  Otherwise, when the rare electric field disappears, the separated charges will quickly recombine after the rare electric field disappears.  The breakage of the wire is evident from Panels D and E of Fig.~\ref{fig:AutoIonization}. It removes the direct pathway to ion recombination.  Average hydrogen-bond lifetimes are of the order of picoseconds, while the destabilizing electric field persist for only tens of femtoseconds.  So, it is unusual for a breaking of a hydrogen bond wire to occur during the specific period of time when the electric field acts on that particular wire.  After the wire is broken, with the charges separated, and rapid recombination is inhibited, the solvated proton (i.e., the hydronium ion) can then diffuse away from its parent hydroxide.  

According to these theoretical results, therefore, auto-ionization is the coincidence of a large but fleeting electric field fluctuation and a switch in hydrogen bond allegiance.  Both are rare events, the former much rarer than the latter, and the two must occur simultaneously---at the same point in space and time.  Figures~\ref{fig:AutoIonization} D~and~E show typical configurations just before and after the system crosses the transition state for this process.  Experimental observations find that the recombination of initially separated hydroxide and hydronium ions is diffusion-limited with a reactive inter-ionic separation of about 8 \AA.  The mechanism exhibited in Fig.~\ref{fig:AutoIonization} explains why this length is significantly larger than the typical separations between nearest-neighbor and second-nearest-neighbor water molecules.  But direct experimental tests of this mechanism for the fundamental kinetic step of pH are not yet available.

\section{Lecture Two: Solvation}

The previous lecture describes the behavior of pure liquid water, stressing how its local structure is tetrahedral, and how its density and polarization fluctuations obey Gaussian statistics to a good approximation.  This second lecture describes how the nature of these fluctuations determines water's behavior as a solvent.  The central quantity to be considered is solvation free energy---the reversible work done on the solvent to accommodate a solute molecule.  This quantity determines the probability of solvation and its associated driving forces.

These forces can be strong, such as when water successfully outcompetes powerful ionic bonds, or when it completely shuns solutes, as happens in oil-water de-mixing.  Amphiphilic substances add geometric frustration to the physics of solvation, where each molecule has two distinct portions at a fixed separation, one that binds water and the other that repels water.   The fixed separation is a geometric constraint that frustrates water structure, which in turn can lead to aggregated nano structures, like membranes and micelles.  Proteins are large amphiphilic molecules for which water significantly affects dynamics and thermodynamics, for instance by providing one of the driving forces for protein folding and assembly.  

This lecture describes underlying principles and simple quantitative estimates of free energies related to these behaviors.  A few consequences are described in the next lecture.

\subsection{Solvation free energies}

\begin{figure}\begin{center}
\includegraphics{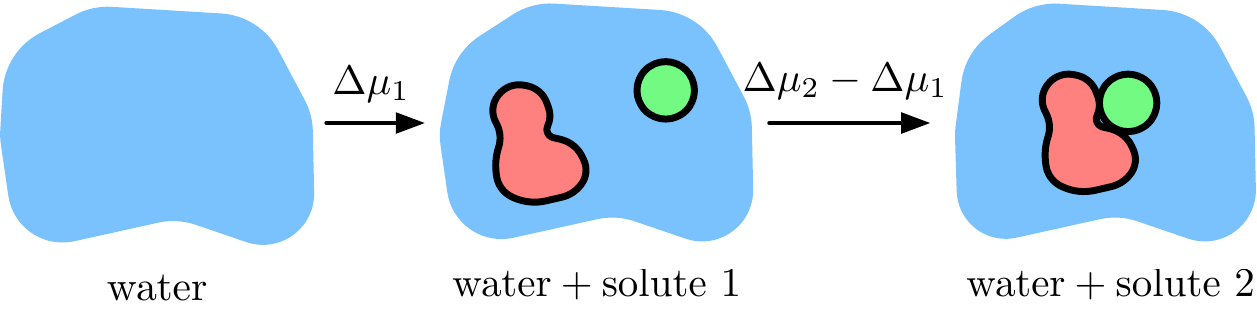}
\end{center}
\caption{\label{fig:SolvationCartoon}Solvation free energies and the forces of assembly.  The excess chemical potentials for the separated pair (the red and green particles in the middle picture) and the complexed pair (the red and green particles in the right-hand picture) are the solvation free energies ~$\Delta\mu_1$ and $\Delta \mu_2$, respectively.  The change in solvation free energy, $\Delta \mu _2 - \Delta \mu _{1}$ is the reversible work done on the liquid to assemble the complexed red-green solute from the separated red-green pair.}
\end{figure}

Imagine a system where initially there is only bulk water and a solute in vacuum far away.  The solute is then slowly inserted into the water (Fig.~\ref{fig:SolvationCartoon}).  This insertion requires performing some reversible work, which is equal to the free energy difference~$\Delta\mu$ between the system before and after insertion.  This free energy can be calculated from statistical mechanics as a ratio of partition functions,
\begin{equation}
e^{-\beta\Delta\mu} = \frac{\int dx\, e^{-\beta E_1(x)}}{\int dx\, e^{-\beta E_0(x)}}\,\,,
\label{eqn:FreeEnergyDefn}
\end{equation}
where $x$ denotes a configuration of the system, $E_0(x)$ is the energy of the pure solvent and $E_1(x)$ is the energy of the solvent when the solute is in the liquid.  The energy function $E_1(x)$ is parameterized by the state of the solute, so changing the condition of the solute generally changes that function.

The quantity $\Delta\mu$ is called the \emph{solvation free energy}.  In a dilute solution, where the solutes do not interact with each other, the chemical potential $\mu$ of the solute is given by
\begin{equation}
\beta\mu = \beta\Delta\mu + \ln(\rho/\rho_0),
\end{equation}
where $\rho$ is the concentration of the solute, relative to its concentration~$\rho_0$ in the standard state.  The logarithm on the right-hand-side is the only contribution to $\beta \mu$ when the solute and the water do not interact.  This logarithm is thus the ideal gas chemical potential.  For this reason, the solvation free energy is sometimes also called the solute's \emph{excess} chemical potential---the excess relative to that of an ideal gas.  One way to deduce excess chemical potentials from experiments is to measure partitioning fractions of solutes in two coexisting fluids.  Connections like that between $\Delta \mu$ and experiments are discussed in standard texts, e.g.~\onlinecite{Chandler1987} and \onlinecite{DillBromberg2010}.

When two or more solutes are dissolved in water, their solvation free energy can depend on their relative positions.  This dependence may be due to packing effects, for example, or the structure of the hydrogen-bonding network of water around one solute being perturbed by the presence of the other solute.  Regardless of how this dependence arises, the solvation free energy of the two solutes when they are nearby ($\Delta \mu _2$ in Fig.~\ref{fig:SolvationCartoon}) may differ from its value when they are far apart ($\Delta\mu_1$ in Fig.~\ref{fig:SolvationCartoon}).  If the difference $\Delta \mu _2 - \Delta \mu_1$ is negative, then the associated state is favored.  If it is positive, on the other hand, the dissociated state is favored.  The solvent provides a force of assembly or disassembly, respectively, depending upon whether $\Delta \mu_2 - \Delta \mu_1$ is negative or positive.  This type of force is what drives oily particles together in water and is ultimately responsible for the hydrophobic effect, the main topic we discuss in Lecture~Three.

The excess chemical potential is but one particular free energy that can be computed for a solute-water system.  In particular, consider a variable $\lambda$ that interpolates between one solute and another.  For instance, $\lambda$ may be the distance from the solute to the liquid-vapor interface, or it may be the distance between two dissolved solutes, or a parameter that is used to progressively ``create'' or ``grow'' solutes into the solvent.  Denote by $\lambda(x)$ the value of~$\lambda$ for a particular configuration~$x$.   At equilibrium, configurations with specific values of $\lambda$ occur with probability
\begin{equation}
P(\lambda) \propto \int \text{d}x\, \delta \left( \lambda(x) - \lambda \right) \exp[-\beta E_1(x; \lambda(x))].
\end{equation}
The right-hand side of this equation is a partition function for configurations constrained to the surface $\lambda(x) = \lambda$. The quantity $-\kB T \ln P(\lambda)$ is therefore the free energy of the system constrained to have that particular value of~$\lambda$.  When $\lambda = 1$ and $\lambda = 0$ correspond to the solvent with and without the solute, respectively, then $- k_{\mathrm{B}} T \ln [P(1)/P(0)] = \Delta \mu$.  At intermediate values of $\lambda$, $- k_{\mathrm{B}} T \ln [P(\lambda)/P(0)]$ serves as a useful generalization of solvation free energy.

We now apply this generalization in two ways, first to estimate solvation free energy of excluded volume (i.e., the reversible work to make space for a solute in a liquid solvent), and then to estimate solvation free energy due to the charge of a solute (i.e., the reversible work to polarize the solvent).

\subsection{Solvation of small excluded volumes}
To compute the solvation free energy for an excluded volume $\V$, we use for $\lambda(x)$ the observable $N(x)$, which denotes the number of water molecules in a volume of that size in the liquid.  In other words, we consider the probability that $N$ molecules exist in a volume $\V$, $P_\V(N)$.  The excess chemical potential for a particle of volume $\V$ absent any other interactions with the solvent is the reversible work to empty that volume, which is to say
\begin{equation}
\label{Pv(0)}
 \Delta \mu_\V = -\kB T \ln P_\V(0).  
 \end{equation}
 Gerhard Hummer, Lawrence Pratt and their co-workers introduced this way of considering solvation~\cite{HummerEtAl1996}, and we follow their lead.
 
Specifically, from the empirical observation discussed in the previous lecture---that density fluctuations in small volumes are Gaussian---we write
\begin{equation}
\label{GaussianDistribution}
\PvN \approx  \frac{1}{\sqrt{2\pi \,\sigma_\V \,}}\exp\Biggl[-\frac{(N - \langle N \rangle_\V)^2}{2 \sigma_\V}  \Biggr],
\end{equation}
where $\langle N \rangle_\V = \rho \, \V$ is the average number of water molecules in the volume~$\V$, $\rho$ being the mean bulk density of the liquid, and
\begin{equation}
\sigma_\V = \langle (\delta N)^2 \rangle _\V
\end{equation}
is the mean-square fluctuation in the number of molecules in that volume.  Accordingly, the solvation free energy of an excluded volume~$\V$ is
\begin{equation}
\Delta\mu_v \, \approx \, \kB T \Bigl[  \frac{\V^2\,\rho^2}{2\sigma_\V} + \frac{1}{2}\ln(2\pi \sigma_\V) \Bigr]\,\,.
\label{eqn:DeltaMuSmall}
\end{equation}

Equation~\eqref{eqn:DeltaMuSmall} is an important result that holds to the extent that the Gaussian distribution, Eq.~\eqref{GaussianDistribution}, is valid.  This in turn requires that $\V$ not be too large.  In particular, water at standard conditions is close to coexistence with its vapor phase, and as such, emptying a large enough volume can nucleate a vapor bubble and its associated liquid-vapor interface.  The Gaussian approximation to $P_\V(N)$ does not include this physics of phase change, but rather includes only the physics of fluctuations that do not move the region within $\V$ far from its typical liquid state.  We will see that this criterion limits the usefulness of Eq.~\eqref{eqn:DeltaMuSmall} to $\V<1$ nm$^3$.    

Keeping to $\V < 1$ nm$^3$, Eq.~\eqref{eqn:DeltaMuSmall} provides a means for estimating solvation free energies associated with excluded volume in water.  The values of $\langle N \rangle_\V$ and $\sigma_\V$ can be estimated simply.  Specifically, the average number of water molecules in~$\V$ is $\langle N \rangle_\V = \rho\,\V$, and the mean-square fluctuations are given by
\begin{equation}
\sigma_\V = \int_{\vr\in \V} \text{d}\vr\, \int_{\vr'\in \V} \text{d}\vr'\, \langle \delta\rho(\vr) \delta\rho(\vr') \rangle\,,
\end{equation}
where $\delta \rho (\vr)$ is the fluctuation (i.e., deviation from the mean) of the molecular density at position $\vr$.  The integrand is related to the pair distribution function $g_{\O\O}(\vr)$,
\begin{equation}
\label{PairCorrelation}
\langle \delta\rho(\vr)\, \delta\rho(\vr') \rangle = \rho \, \delta(\vr-\vr') + \rho^2 [g_{\O\O}(|\vr - \vr'|) - 1],
\end{equation}
and can thus be calculated when $g_{\O\O}(r)$ is known.  Solvation free energies estimated from these formulas agree well with those obtained from computer simulations, and these formulas form a convenient basis for accurate estimates of hydration free energies for small oily molecules in water.  The extent of accuracy is illustrated later in this lecture with Fig.~\ref{fig:SolvationScaling} below.

\begin{figure}\begin{center}
\includegraphics{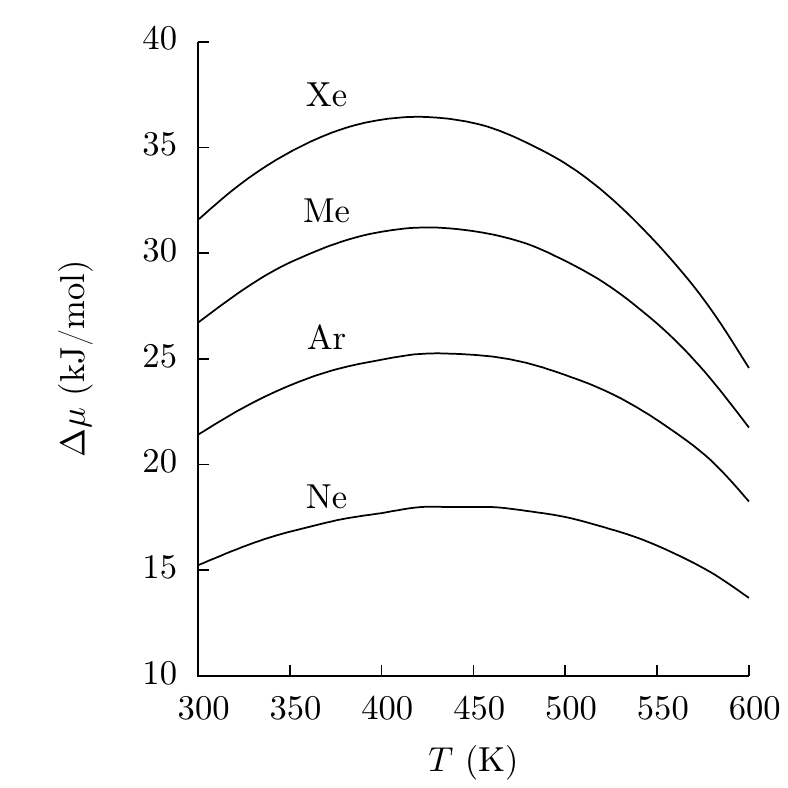}
\end{center}
\caption{\label{fig:EntropyConvergence} Predictions from Eq.~\eqref{eqn:DeltaMuSmall} of solvation free energies for hard spheres in water, with hard-sphere radii comparable to those of Ne, Ar, Me (methane) and Xe, as.  Around $T\approx 400\,$K, all curves have the same slope (i.e., the same solvation entropy).  Adapted from Ref.~\onlinecite{GardeEtAl1996}.}
\end{figure}

Equation~\eqref{eqn:DeltaMuSmall} shows that the free energy to exclude small volumes is primarily entropic.  In particular, this free energy grows with increasing temperature.  To the extent that excluded volume effects are dominant, therefore, small solutes become \emph{less} soluble, not more, as temperature increases.  This trend is indeed observed in experimentally measured solubilities of small apolar molecules in water.  A trove of such data is found in Charles Tanford's classic monograph~\cite{Tanford1973}.  At temperatures above about~$50^\circ$\,C, however, the structure of water begins to change substantially (Fig.~\ref{fig:XRaysGofR}), so that the solvation free energy of a small solute stops being proportional to temperature.  The theoretical predictions are graphed in Fig.~\ref{fig:EntropyConvergence}.

To gain further insight, consider $\V$ significantly larger than a correlation volume.  Such sizes are at the upper limit of where the Gaussian approximation is accurate.  But for such volumes, the compressibility theorem mentioned in the previous lecture gives
\begin{equation}
(\rho\,\V)^{-1}\,\sigma_v = \frac{\langle (\delta N)^2 \rangle_\V}{\langle N \rangle_\V} \approx \frac{\partial \rho}{\partial \beta p} \equiv \chi \,.
\end{equation}
An approximate expression for the solvation free energy is thus
\begin{equation}
\Delta\mu \approx \kB T \Biggl[ \frac{\rho\V}{2\chi} + \frac12\ln(2\pi \rho\V \chi) \Biggr].
\end{equation}
This simple estimate is qualitatively correct, but overestimates the more generally accurate Eq.~\eqref{eqn:DeltaMuSmall}. Nevertheless, it immediately shows that hydration free energies due to excluded volume of small apolar species are typically $5$~to~$10\,\kB T$ because the quantity $\chi$ has a value of about $0.06$ for water for a range of temperatures around ambient conditions.  It further illustrates that the solvation free energy grows roughly linearly with increasing volume in the regime where the Gaussian approximation is valid.  

\begin{figure}\begin{center}
\includegraphics{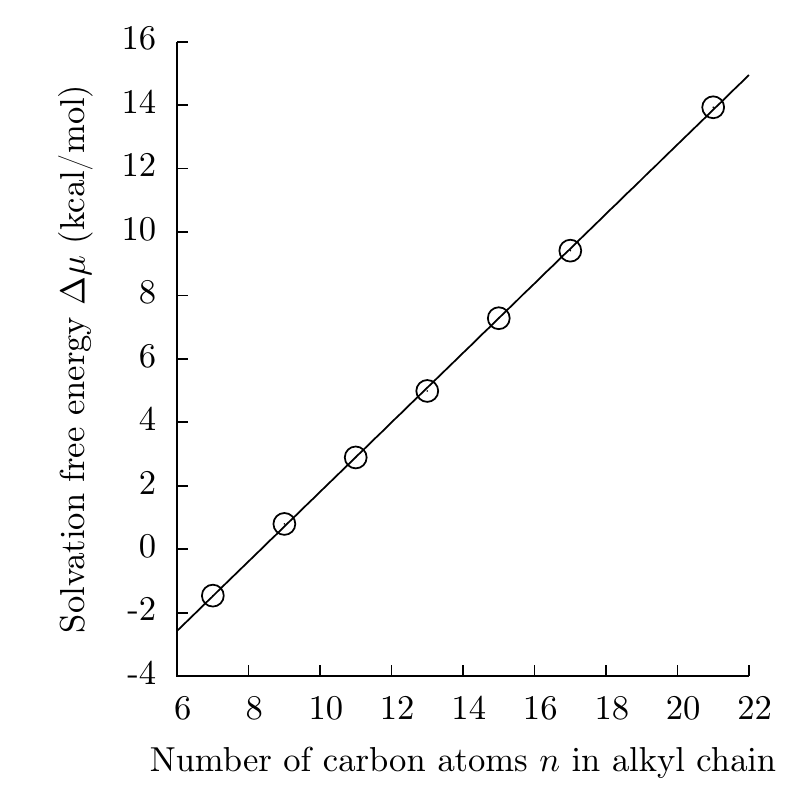}
\end{center}
\caption{\label{fig:TanfordAlkanes}Solvation free energies of $\C\H_3 (\C\H_2)_{n-1} \C\O\O\H$ as a function of chain length.  The line is a linear fit to the data.  Adapted from Ref.~\onlinecite{SmithTanford1973}.}
\end{figure}

Figure~\ref{fig:TanfordAlkanes} shows how the experimentally measured solvation free energies of small-to-medium alkyl chains scale with chain length.  These solutes are extended but thin, so they can be treated with the theory we described above for small solutes.  As chain length and solvent-excluded volume are proportional, these results illustrate that solvation free energies of small solutes are proportional to excluded volume.  In many discussions, e.g., Ref.~\onlinecite{StillEtAl1990}, these same results have been used to argue that solvation free energies of even small solutes may be modeled as scaling with solvent-accessible surface area (SASA), because chain length and SASA are also proportional.  However, the fact that the apparent surface tension that would result is much smaller than the measured water-oil surface tension, and that the alkyl chains become less soluble, not more, as temperature is increased, shows that the proportionality with SASA is a misleading coincidence.  References~\cite{Chandler2005} and~\cite{Tanford1979} provide further discussion on this point.

Equation~\eqref{eqn:DeltaMuSmall} rationalizes a phenomenon called \emph{entropy convergence}.  To discuss it, we introduce the entropy of solvation, which is the entropic component of the solvation free energy, given by
\begin{equation}
\Delta s= -\frac{\partial \Delta\mu}{\partial T}.
\end{equation}
It is found that along the saturation curve of water, the entropy of solvation of many different small solutes is different, but surprisingly, converges to a common, small value at a temperature of about~$400\,$K.  This convergence is illustrated in Fig.~\ref{fig:EntropyConvergence}.  Around this temperature, it is found that the factor $\sigma_\V$ is nearly athermal, so the solvation free energies are essentially proportional to $T \rho^2(T)$.  This combination is non-monotonic with temperature, and is maximal around $T=400\,$K.  Thus, its derivative is zero there, leading to a vanishing entropy of solvation at that temperature.  The observed small but non-zero value arises from the second term of Eq.~\eqref{eqn:DeltaMuSmall} and the small temperature dependence of $\sigma_\V$.

A similar convergence is also found around $T=400\,$K in the per-residue entropies of unfolding of proteins.  Though it superficially resembles the entropy convergence of small solutes, the origin of this effect is actually more complicated and due to the interplay of small and large length-scale physics that we discuss in general at the end of this lecture and in Lecture~Three.  Reference~\onlinecite{HuangChandler2000a} provides further discussion on this particular point.

\subsection{Solvation of ions}

So far, we have discussed solvation of excluded volume.  While all solutes exclude water from some region of space, most solutes act with additional forces on water, including weak dispersive attractions and electrostatic interactions.  Weak interactions change the solvation free energy quantitatively but not qualitatively.  On the other hand, owing to the large dielectric constant of water ($\epsilon \approx 80$), electrostatic interactions are significant, and often dominate solvation behavior when they are present.  For instance, since we know that water dissolves many salts, it must be the case that the solvation free energies of the separate ions is larger than the strength of the ionic bonds between them in a crystal, which is about $1\,$eV, or about $40\,\kB T$ at ambient conditions.

\begin{figure}
\hfill\begin{tabular}{cc}
\includegraphics{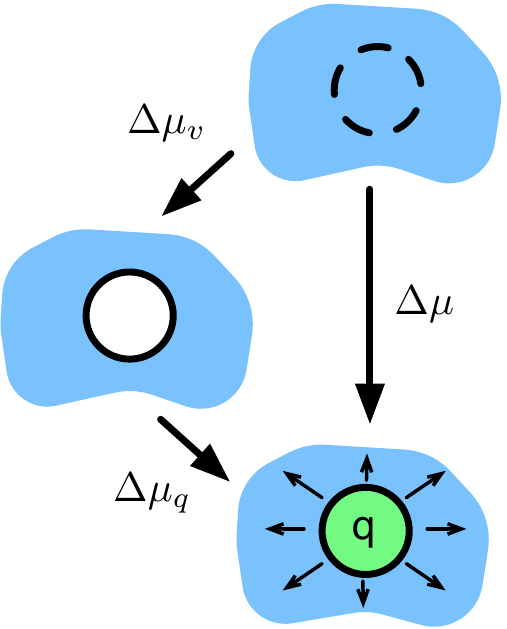}&
\includegraphics{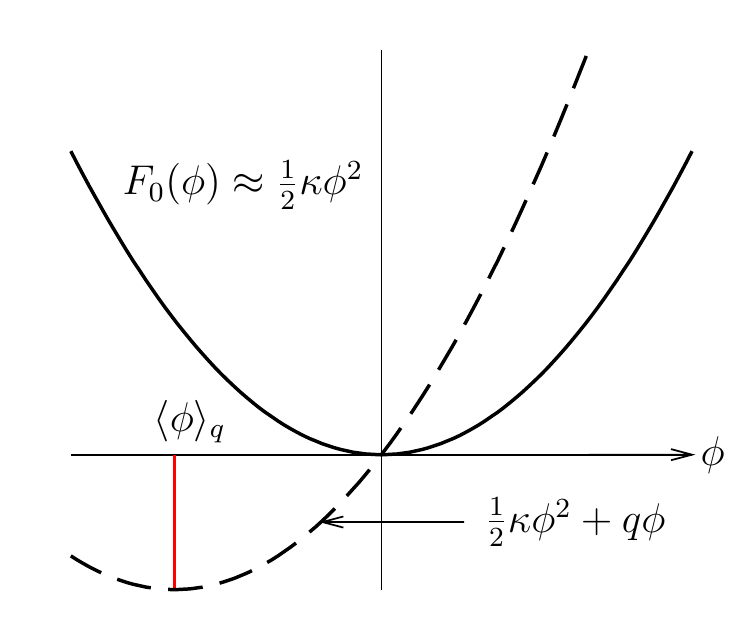}\\
(a)&(b)
\end{tabular}\hfill
\caption{\label{fig:SolvatingIonCartoon}(a) Calculating the solvation free energy of an ion in two steps.  A free energy~$\Delta\mu_v$ is needed to expel waters from the ion's hard core.  A free energy~$\Delta\mu_q$ is then required to place a charge~$q$ in the ion and thus polarize the solvent.  (b) Free energies of the potential~$\phi$ at the center of an ion, before (solid) and after (dashes) the charge is inserted.  The induced polarization of the solvent shifts the mean potential to $\langle\phi\rangle_q = -q/\kappa$.}
\end{figure}

As a model for electrostatic interactions, we estimate the solvation free energy of a single ion, as outlined in Fig.~\ref{fig:SolvatingIonCartoon}(a).  An ion typically interacts with water by excluding a water from its hard core and by electrostatically polarizing the surrounding solvent.  To be concrete, a Cl$^-$ ion excludes water from a sphere with radius of about $2.5\,$\AA, and can be modeled as having a point charge of $-e$ at its center.  The free energy~$\Delta\mu_v$ to expel the water from the hard core is given by Eq.~\eqref{eqn:DeltaMuSmall}.  The free energy~$\Delta\mu_q$ to then place a point charge~$q$ at the ion is given by
\begin{equation}
e^{-\beta\Delta\mu_q} = \frac{\int \text{d}\phi\, e^{-\beta[F_0(\phi) +q\phi]}}{\int \text{d}\phi\, e^{-\beta F_0(\phi)}},
\label{eqn:DeltaMuChargeSetup}
\end{equation}
where $\phi$ is the value of the electrostatic potential at the center of the ion given that a cavity for it has been created, and $F_0(\phi)$ is the free energy associated with a particular value of this potential, equal to $-\kB T \ln P(\phi)$.  As outlined previously, the statistics of electric field and potentials in water are Gaussian, so $F_0(\phi)$ is quadratic in $\phi$.  Moreover, since the water is charge neutral, the mean value of $\phi$ should be essentially zero\footnote{It wont be exactly zero, however, in the presence of large enough solutes that cause significant inhomogeneity.  Liquid-vapor interfaces have a small surface dipole moment density, since a small number of water molecules at the interface tend to align their dipoles with the surface normal.  The bulk liquid has been measured to be at a potential of about $0.1\,$V higher than the bulk vapor.  This potential difference is the so-called \emph{zeta potential} of the interface.  For a discussion of these effects and how the presence of ions modifies the structure of the liquid-vapor interface, see Ref.~\onlinecite{PetersenSaykally2006}.}.  Hence, we approximate
\begin{equation}
F_0(\phi) \approx \frac12 \kappa \phi^2,
\end{equation}
where $\kappa$ is related to the fluctuations of~$\phi$ in the absence of the charge by equipartition,
\begin{equation}
\langle (\delta\phi)^2 \rangle = \frac{\kB T}{\kappa}.
\end{equation}
These fluctuations are, in turn, controlled by the dielectric constant of water and the size of the ion, as discussed below.  With this expression for $F_0(\phi)$, we can evaluate the right-hand side of Eq.~\eqref{eqn:DeltaMuChargeSetup} and simplify the result to obtain
\begin{equation}
\Delta\mu_q = -\frac12 \beta q^2 \langle (\delta\phi)^2 \rangle = -\frac{q^2}{2\kappa}.
\end{equation}

When the charge is placed into the ion, the surrounding solvent is polarized.  As depicted in Fig.~\ref{fig:SolvatingIonCartoon}(b), the potential fluctuations are unchanged but the mean value of the potential shifts to
\begin{equation}
\langle\phi\rangle_q = -q/\kappa
\end{equation}
This induced potential acts as a \emph{reaction field} of the solvent to the presence of the charge.
Polarizing the solvent in this way has a free energy cost of
\begin{equation}
F_0(\langle\phi\rangle_q) = \frac12 \kappa (q /\kappa)^2 = \frac{q^2}{2\kappa},
\end{equation}
but results in a favorable interaction energy of
\begin{equation}
q \langle\phi\rangle_q = -\frac{q^2}{\kappa}
\end{equation}
between the ion and the polarized solvent.

The value of~$\kappa$ can be determined by relating the fluctuations in the dipole moment density field of water to its dielectric constant and the geometry of the cavity (for details, see Ref.~\onlinecite{SongChandlerMarcus1996}).  It is found to be
\begin{equation}
\kappa = R \Bigl( 1 - \frac1\epsilon \Bigr)^{-1},
\end{equation}
where $R$ is the radius of the ion.  This result is identical to the continuum electrostatics result~\cite{Griffiths1999}, where the work required to move a charge $q$ from vacuum into a spherical cavity of radius~$R$ carved into a dielectric material with dielectric constant~$\epsilon$ is
\begin{equation}
\Delta\mu_q = -\frac{q^2}{2 R} \Bigl( 1 - \frac1\epsilon \Bigr).
\label{eqn:Born}
\end{equation}
Equation~\eqref{eqn:Born} is the Born solvation formula.  It is negative, so it is always favorable to solvate an ion.  It scales as $q^2$, so divalent ions are much more strongly solvated than monovalent ones.  Finally, it scales as $1/R$, so smaller ions are more soluble than larger ones.

\begin{figure}\begin{center}
\includegraphics{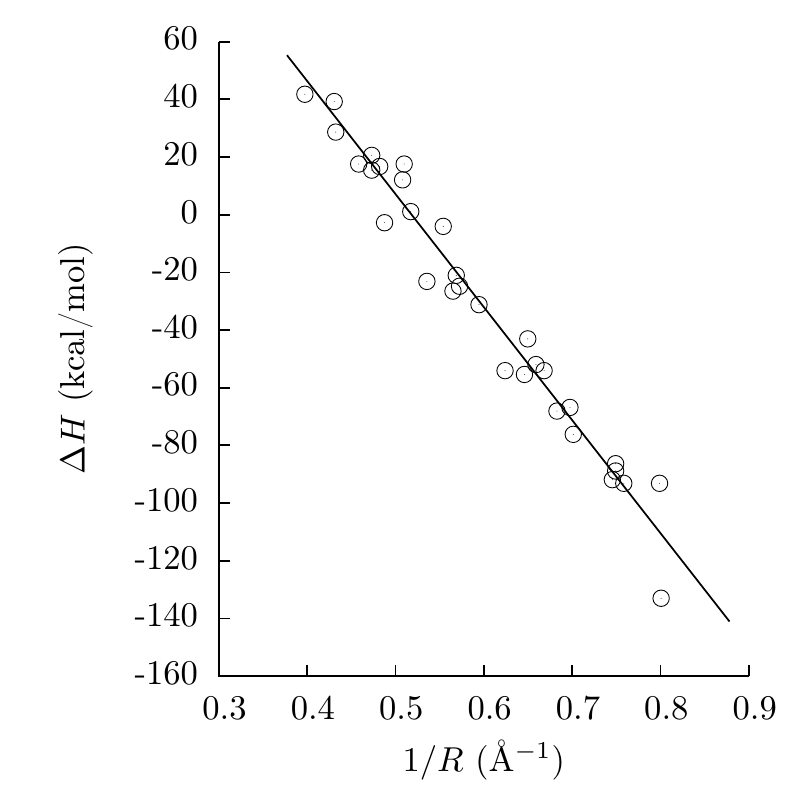}
\end{center}e
\caption{\label{fig:TestOfBorn}Enthalpies of hydration ($\Delta\mu - T [\partial\Delta\mu/\partial T]$) for a variety of monovalent ions as a function of inverse ionic radius, as measured experimentally (circles) and predicted by Eq.~\eqref{eqn:Born} (line).  Adapted from Ref.~\onlinecite{RashinHonig1985}.  Similar correlations are seen for polyvalent ions.}
\end{figure}

For water, $\epsilon$ is so high that the term in brackets is virtually unity.  To within an order of magnitude, the solvation free energies of ions are about $e^2 / (1\,\text{\AA}) \approx 10\,\text{eV} \approx 400\,\kB T$.  As anticipated, these free energies are indeed sufficiently large to overcome the ionic bonds in salt crystals.  Figure~\ref{fig:TestOfBorn} shows a more detailed comparison of the Born solvation formula to experimental enthalpies of solvation.  It is clear that the Born solvation formula works quite well for all of these cases, and thus captures the dominant physics of solvation for ions.

\subsection{Solvation of large solutes}
\label{sec:SolvationLarge}

In both of the scenarios considered above, we could calculate solvation free energies from the statistics of small structural fluctuations in water.  The solute, whether ideal or charged, acted as a small perturbation that did not fundamentally change the nature of the liquid.  As solutes increase in size beyond about $1\,$nm, however, water responds to their presence in a massively collective fashion, undergoing nano-scale reorganization manifesting a macroscopic phase transition---the liquid-vapor phase transition---that occurs at thermodynamic conditions very close to those of ambient water.  The solvation free energy is then dominated by the surface energy of the interface between the two phases, which for macroscopic solutes is given by
\begin{equation}
\Delta\mu \to \gamma A, \qquad\text{(large solutes)},
\label{eqn:DeltaMuLarge}
\end{equation}
Here, $\gamma$ is the surface tension between water and the solute (about $72\,$mJ/m$^2$ for a solute that only excludes volume) and $A$ is the surface area of the solute.

\begin{figure}\begin{center}
\includegraphics{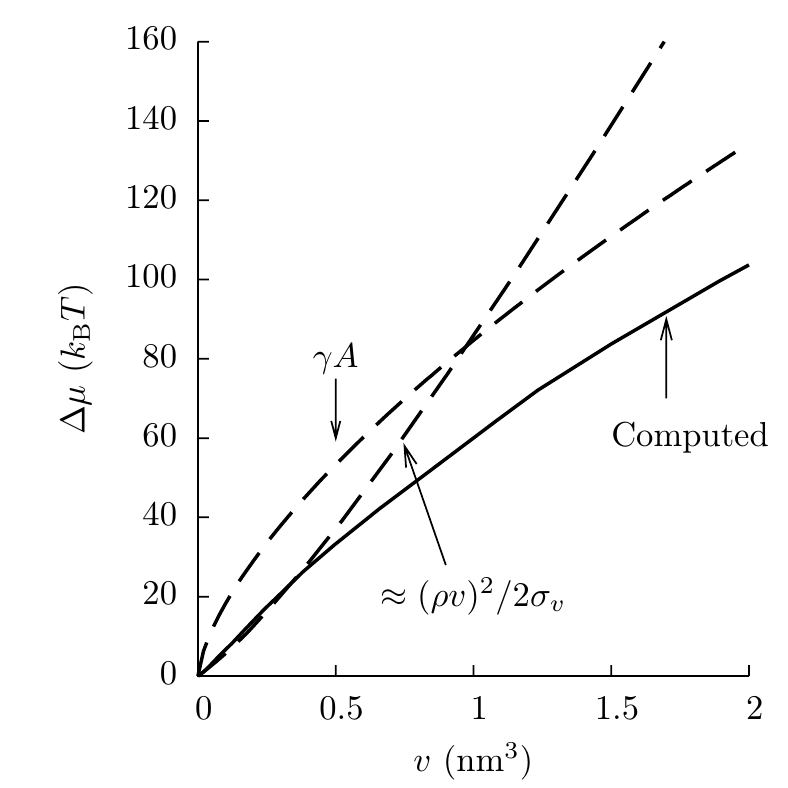}
\end{center}
\caption{\label{fig:SolvationScaling}Solvation free energies for hard spheres of volume~$v$, computed through direct simulation with the SPC/E model of water at ambient conditions (black, Ref.~\onlinecite{HuangGeisslerChandler2001}).  The small length-scale estimate (Eq.~\eqref{eqn:DeltaMuSmall}) and the large length-scale estimate (Eq.~\eqref{eqn:DeltaMuLarge}) for the same model are also shown.}
\end{figure}

For nanometer-sized solutes, there is a gradual crossover between the small length-scale behavior described by Eq.~\eqref{eqn:DeltaMuSmall} and the large length-scale behavior described by Eq.~\eqref{eqn:DeltaMuLarge}.  This crossover is depicted schematically in Fig.~\ref{fig:SolvationScaling} for hard sphere solutes of volume~$v$.  The two estimates for $\Delta\mu$ intersect around $v$ of $1\,$nm$^3$.  This sets the scale for where the crossover is relevant, though corrections to Eq.\eqref{eqn:DeltaMuSmall} are already important for $v \gtrsim 0.5\,$nm$^3$.  Solvation free energies eventually do tend to $\gamma A$, but convergence to this limit is slow because the ratio of these two quantities tends to~$1$ only as $v^{-1/3}$.  Corrections to Eq.~\eqref{eqn:DeltaMuLarge} are significant even for excluded volumes measuring tens of cubic nanometers.

\begin{figure}\begin{center}
\includegraphics{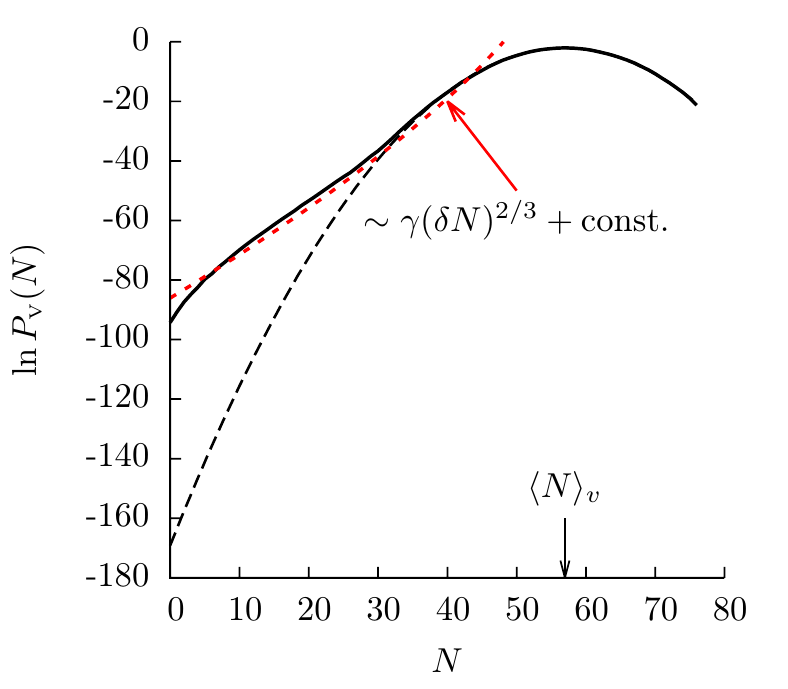}
\end{center}
\caption{\label{fig:PvNFatTails}Water number distribution, $\PvN$, for a cube of volume~$v = 1.7\,$nm$^3$ computed from importance sampling with a computer simulation of the SPC/E model of water at ambient conditions (solid), and the distribution that would be expected if the statistics of these number fluctuations were Gaussian (dashed).  The fat tail is an indication of the crossover in solvation free energies to the large-length scale regime.  Adapted from Ref.~\onlinecite{PatelVarillyChandler2010}.}
\end{figure}

The breakdown of Eq.~\eqref{eqn:DeltaMuSmall} for larger~$v$ indicates that the $\PvN$ distributions for these larger volumes deviate significantly from Gaussian behavior.  This is indeed the case.  Figure~\ref{fig:PvNFatTails} shows an explicit calculation of $\PvN$ where $\V$ is a cube of volume~$1.7\,$nm$^3$.  The breakdown of Gaussian statistics appears as a fat tail in this distribution.  This fat tail indicates that there is a mechanism beyond Gaussian density fluctuations that can be involved when removing $N$~waters from the probe volume~$\V$.  The fat tail in $\ln \PvN$ is well-described by a function that scales as $-\gamma (\delta N)^{2/3}$, which describes the free energy to form an empty cavity of volume~$\delta N / \rho$ inside of the probe volume.  The physical consequences of this fat tail and corresponding crossover in length-scale are discussed in the next lecture.

\section{Lecture Three: Hydrophobicity and Self-Assembly}

This lecture is about hydrophobicity, which essentially means it is about how and why oily species---termed hydrophobic species---separate from water in nanometer scale structures.  ``Hydrophobic'' is actually a misnomer, because the underlying physics is not about oil fearing water, but rather about hydrogen bonds that are lost when water mixes with oil.  A hydrophobic species is simply a molecule or complex of molecules that binds more weakly to water than water binds to itself.  A hard sphere is the simplest (and idealized) example.  Surfaces of nominally hydrophilic molecules can also be hydrophobic when the geometry of possible water-surface binding sites is incommensurate with favorable hydrogen patterns of liquid water. 

Hydrophobic forces are the water-mediated interactions between oily species in water that cause these species to segregate or demix from water.  Macroscopic demixing of oil and water manifests a first-order phase transition.  Such manifestations require clusters of size larger than a critical nucleus.  A small enough hard sphere in water will not trigger the physics of phase separation.  Hydrophobic forces therefore become significant only for sufficiently large hydrophobic species.  We will see that ``sufficiently large'' implies hydrophobic surfaces of low curvature that extend over lengths of the order of 1 nm or more.         

Lecture Two provides the background to these ideas.  Indeed, the $1\,$nm scale mentioned here is related to a finding discussed in the previous lecture, where we describe how solvation free energies are related to statistics of solvent density fluctuations and solvent polarization fluctuations.  We show there that these statistics are essentially Gaussian for cases where the solute does not largely perturb the liquid.  But we also show there that this situation changes in the disruptive presence of sufficiently extended hydrophobic surfaces.  This lecture focuses on the consequences of these changes and a theory that embodies the underlying physics.  Ref.~\cite{Chandler2005} is review of the topic written with the same perspective as that of this lecture.

\begin{figure}
\begin{center}\includegraphics{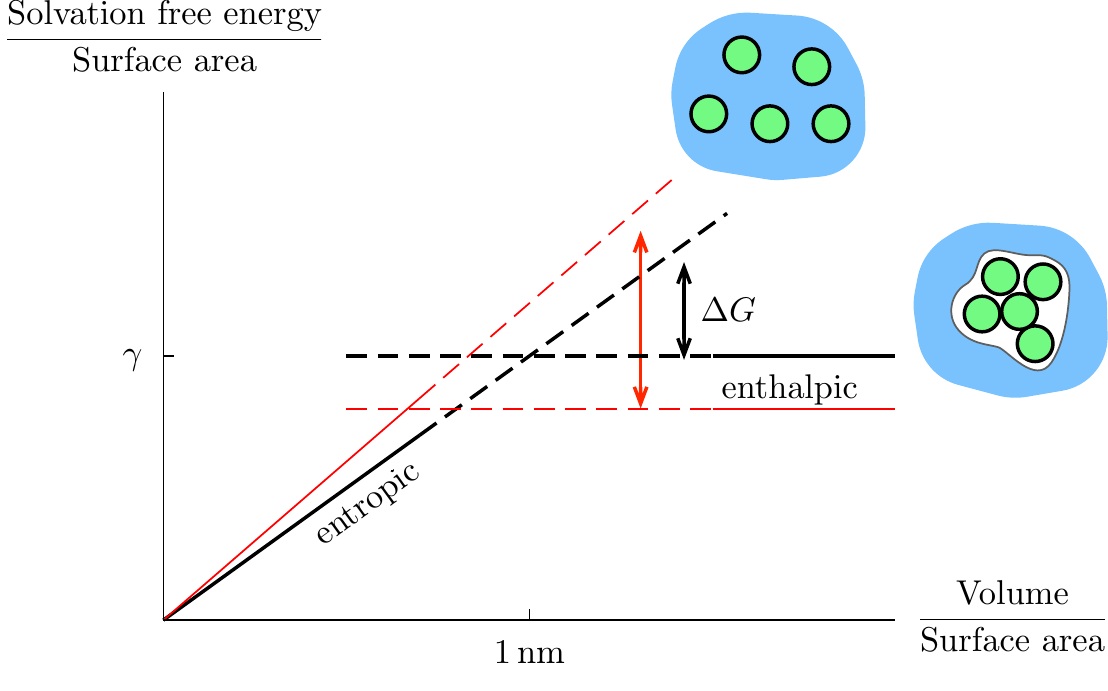}\end{center}
\caption{\label{fig:HydrophobicAssembly}Driving force~$\Delta G$ of hydrophobic assembly.  At room temperature (thick black lines), the crossover length scale is about~$1\,$nm.  At higher temperatures (thin red lines), the crossover length scale decreases and the driving force increases.  After Ref.~\cite{Chandler2005}.}
\end{figure}

\subsection{The driving force for hydrophobic assembly}

Figure~\ref{fig:HydrophobicAssembly} helps to show how the assembly of hydrophobic aggregates in water is explained in terms of the length-scale dependence of solvation free energies discussed at the end of Lecture~Two.  As schematized in that figure, the solvation free energy to solvate $N$ small, dissociated solutes scales as~$N$, and is mostly entropic in origin.  The hydrogen-bonding network of water is stretched but not disrupted by these solutes.  Except for small packing effects common to all dense liquids, the total solvation free energy of this collection of solutes is independent of their positions.  As such, entropy disfavors aggregation, so the solutes disperse evenly throughout the solvent.

If enough of these solutes associate spontaneously, however, it becomes impossible for water to wrap its hydrogen-bonding network around the cluster.  It then becomes preferable to sacrifice some hydrogen bonds completely and nucleate a vapor-liquid interface around the cluster.  The cost of maintaining this interface is primarily enthalpic, and scales as $N^{2/3}$.  Hence, for large enough $N$, it is energetically favorable for small solutes to associate into a larger cluster, with the free energy difference~$\Delta G$ between the dissociated and associated states termed the ``driving force'' for hydrophobic assembly.  At ambient conditions, the critical nucleus that needs to form before aggregation becomes favorable is typically about $1\,$nm in size.

The temperature dependence of the driving force for assembly is non-trivial.  As evidenced by Eq.~\eqref{eqn:DeltaMuSmall}, the solvation free energy of very small solutes is proportional to temperature, because of its entropic origin.  Conversely, the surface tension of most liquids, including water, decreases with temperature.  In other words, whereas rises in temperature make small solutes less soluble, they make large solutes more soluble.  This curious dichotomy in the temperature dependence of solvation free energy has two consequences.  First, the size of the critical nucleus for aggregation is reduced at higher temperatures.  Second, the driving force towards hydrophobic association is stronger in hot water than in cold water.  These consequences are pertinent to protein thermodynamics, where hydrophobic interactions stabilize the core of most globular proteins.  At colder temperatures, the driving force for this hydrophobic collapse decreases, so at low enough temperatures, proteins undergo cold denaturation.

At high pressures, the dominant component to solvating large solutes is not the cost of forming a liquid-vapor interface, but the work needed to create a vacuum bubble in a high-pressure environment, which scales as~$N$.  Hence, above pressures of about $500$\,atm, the driving force for hydrophobic assembly disappears, and proteins undergo pressure denaturation.

\subsection{Micelle Assembly}
\label{sec:micelles}

When small amphiphilic molecules, such as the fatty acids that compose cell membranes, are dissolved in water at low concentrations, entropy favors dispersing them uniformly as monomers.  At higher concentrations, however, the driving force for hydrophobic assembly overcomes this entropic effect, so collections of these molecules assemble into nano-scale clusters, which are called ``micelles.''  In this way, the hydrophobic tails of the amphiphiles are separated from water by a layer of hydrophilic head groups.  Experimentally, the change in aggregation behavior occurs abruptly at a critical micelle concentration, $\rhocmc$, whose non-trivial temperature dependence can be understood from the considerations we have been describing---the competition between small length-scale solvation and interface formation.

\begin{figure}
\begin{center}\begin{tabular}{cc}
\includegraphics{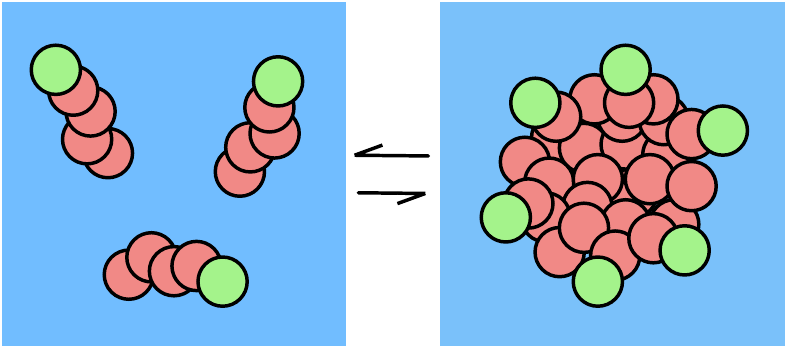}&
\includegraphics{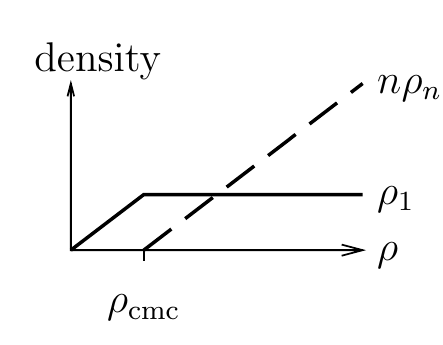}\\
(a)&(b)
\end{tabular}\end{center}
\caption{\label{fig:MicellesIntro} Micelle assembly.  (a) Small amphiphiles dispersed uniformly as monomers in solution (three of which are shown); these monomers are in equilibrium with $n$-mers -- clusters of $n$ amphiphiles (one of which is shown). (b) For amphiphile concentrations~$\rho$ below~$\rhocmc$, only the monomeric species (solid) is present. Above~$\rhocmc$, $n$-mers form (dashes), and adding more amphiphiles only increases the number of $n$-mers}
\end{figure}
 
A simplified model of micelle assembly is guided by Fig.~\ref{fig:MicellesIntro}.  For simplicity, we shall assume that the micelles that form are all the same size, each containing $n$ amphiphiles, with that number $n$ to be determined.  (For large enough $n$ we expect the dispersion of $n$-mer size, $\Delta n$, to be small as a consequence of a law of large numbers, specifically $\Delta n \sim \sqrt{n}$\,.)  Accordingly, the total amphiphile concentration is
\begin{equation}
\rho = \rho_1 + n \rho_n,
\end{equation}
where $\rho_1$ is the density of monomers and $\rho_n$ is the density of $n$-mers.  The law of mass action relates these two quantities by
\begin{equation}
\rho_n = \rho_1^n \exp( -\beta \Delta G ),
\label{eqn:MicelleMassAction}
\end{equation}
where $\Delta G$ is the driving force for assembly of the micelle, namely the free energy difference between one $n$-mer and $n$~monomers.

\begin{figure}
\begin{center}
\includegraphics[width=4.2in]{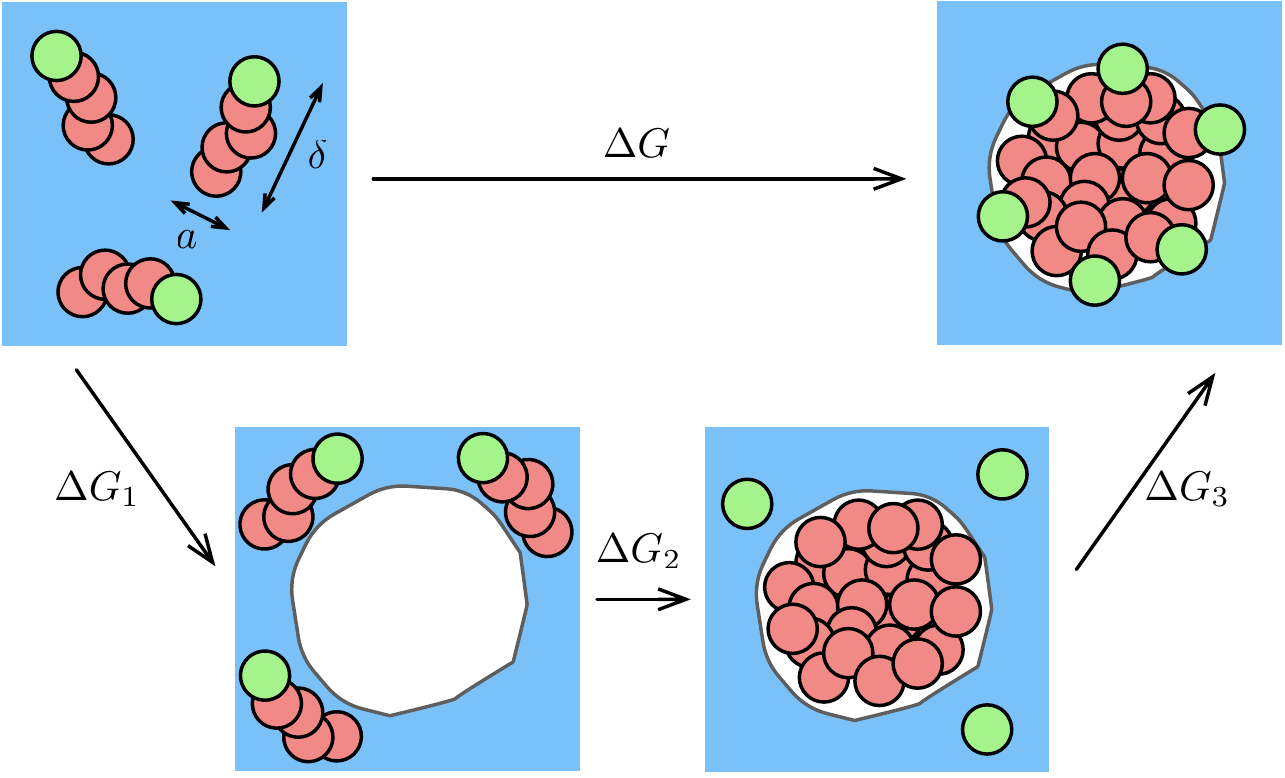}
\caption{\label{fig:MicellesDeltaG}Thermodynamic cycle of micelle formation (adapted from Ref.~\onlinecite{MaibaumDinnerChandler2004}).  The free energy of micelle assembly~$\Delta G$ consists of three contributions: a term~$\Delta G_1$ for forming a cavity of the right size in the solvent, a term~$\Delta G_2$ for detaching the head and tail groups and transferring the hydrophobic tails from solution into the cavity, and a term~$\Delta G_3$ for reattaching the head groups to the tails.}
\end{center}
\end{figure}

To estimate the driving force for assembling an $n$-mer, we use the thermodynamic cycle depicted in Fig.~\ref{fig:MicellesDeltaG}.  The width of the amphiphile is denoted by~$a$ and its typical length is denoted by~$\delta$.  The $n$-mers have radius~$R$ and surface area~$A$.  The assembly is achieved in three steps:
\begin{enumerate}
\item A cavity of the size of a micelle is formed in the solvent.  Since this cavity is large, the free energy to create it is proportional to its area~$A$, so
\begin{equation}
\Delta G_1 \approx \gamma_{\text{water-vapor}} A.
\end{equation}
In terms of $n$, $a$~and~$\delta$, the area~$A$ scales as $a^2 (\delta/a)^{2/3} n^{2/3}$.
\item The head and tail groups are detached, at energy cost $\epsilon_{\text{bond}}$ per molecule, and the tails are transferred from the solution to what will be an oily region.  This is the free energy of transfer~$\Delta\mu_{\text{transfer}}$ of the tails from water to oil, and depends on the tail's size and chemical character.  Finally, what was a water-vapor interface after Step~1 is now a water-oil interface.  Due to dispersive interactions, its surface tension is thus reduced by an amount~$\Delta\gamma$ given by~$\gamma_{\text{water-vapor}} - \gamma_{\text{oil-water}}$.  Hence,
\begin{equation}
\Delta G_2 \approx n \epsilon_{\text{bond}} - n \Delta \mu_{\text{transfer}} - \Delta \gamma A.
\end{equation}
\item Finally, the head groups are reattached to the tails, for an energy gain of $\epsilon_{\text{bond}}$ per molecule.  The additional entropic cost of restricting the $n$ head groups to be at the surface of the micelle is quite large.  Since this restriction is exactly analogous to the requirement of charge neutrality in a polarized dielectric, it can be estimated accurately by analogy~\cite{deGennes1979,Stillinger1983}.  Specifically, the entropic cost of maintaining a separation of a head group from a tail group at a distance~$r$ goes as $1/r$, which makes the effects of this cost isomorphic to that of an electrostatic cost.  Accordingly, the total entropic cost has the same $n^2/R$ functional dependence as the Born solvation energy, Eq.~\eqref{eqn:Born}, where $n$ plays the role of charge and $R \sim n^{1/3}$ is the micelle radius.  The pre-factor is related to the aspect ratio of the amphiphile.  We thus obtain the estimate
\begin{equation}
\Delta G_3 \approx - n \epsilon_{\text{bond}} + \kB T (a / \delta)^{4/3} n^{5/3}.
\end{equation}
\end{enumerate}

\begin{figure}
\begin{center}\begin{tabular}{ccc}
\includegraphics{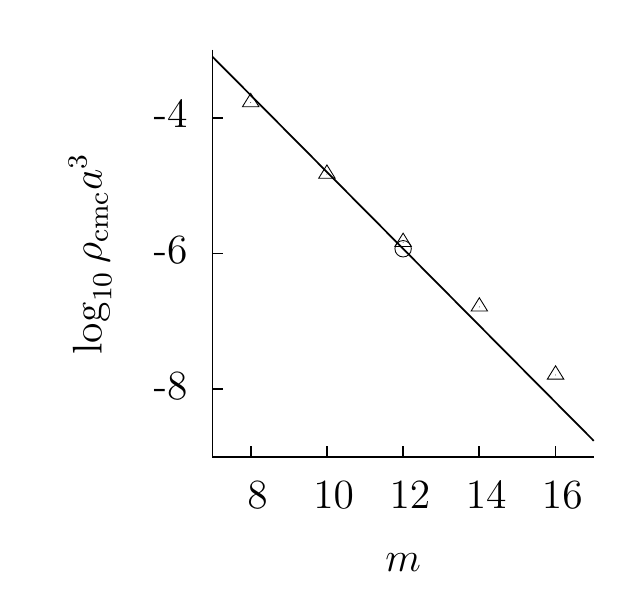}&
\includegraphics{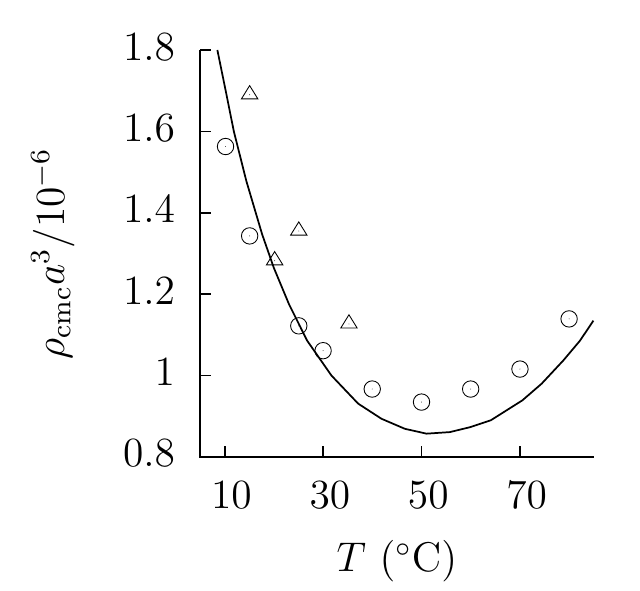}\\
(a)&(b)
\end{tabular}\end{center}
\caption{\label{fig:MicelleResults}Comparison of Eq.~\eqref{eqn:rhocmc} with experimental $\rhocmc$ of $\C\H_3 (\C\H_2)_{(m-1)} (\O\C\H_2\C\H_2)_6 \O\H$.  The single adjustable parameter~$a \approx 3\,$\AA\ is fit to the $m=12$ result at $T = 25^\circ$C.  (a) Dependence of $\rhocmc$ on amphiphile length at $T = 25^\circ$C.  (b) Dependence of $\rhocmc$ on temperature for $m=12$.  In both plots, the data shown is experimental (circles and triangles, from two different experiments) and the result of Eq.~\eqref{eqn:rhocmc} (solid).  Adapted from Ref.~\onlinecite{MaibaumDinnerChandler2004}}
\end{figure}

Above, we focused on the functional dependences of the free energies on $n$, $a$~and~$\delta$.  Adding up the contributions, we obtain the total free energy of assembly:
\begin{equation}
\Delta G \approx \gamma_{\text{oil-water}} a^2 (\delta / a)^{2/3} n^{2/3} - n \Delta \mu_{\text{transfer}} + \kB T (a / \delta)^{4/3} n^{5/3}.
\label{eqn:MicelleDeltaG}
\end{equation}
For a given set of parameters in the above equation, we find the number of monomers $n = n^*$ that minimizes~$\Delta G / n$ to obtain the typical micelle size, and $\rhocmc$ can then be determined through the law of mass action.  We omit the algebra and simply give the final result~\cite{MaibaumDinnerChandler2004}:
\begin{equation}
\ln \rhocmc a^3 \approx c \, (\beta \gamma_{\text{oil-water}} a^2)^{2/3} - \beta \Delta\mu_{\text{transfer}},
\label{eqn:rhocmc}
\end{equation}
with $c =(5832/49)^{1/3} \approx 4.9$.  Figure~\ref{fig:MicelleResults} shows how this prediction compares to the experimental measurements.  The data can clearly be explained quantitatively by invoking only the length-scale dependence of the hydrophobic effect and the geometric constraint that head and tail groups be adjacent.

\subsection{Dewetting transitions in hydrophobic assembly}

As discussed in Section~\ref{sec:SolvationLarge}, and as illustrated in Figs.~\ref{fig:SolvationScaling}~and~\ref{fig:PvNFatTails}, the length-scale dependence of solvation free energies is intimately connected to the formation of liquid-vapor-like interfaces around large solutes.

\begin{figure}
\begin{center}\begin{tabular}{ccc}
\includegraphics{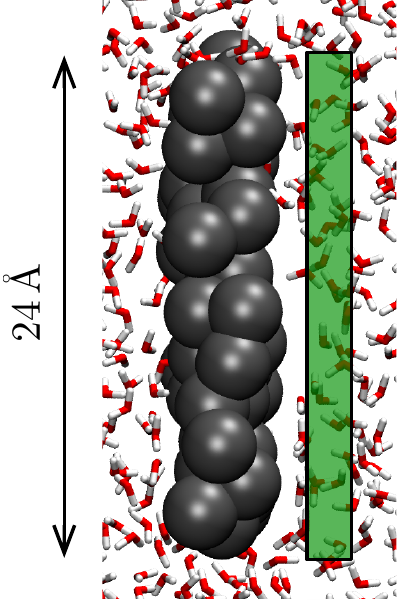}&
\includegraphics{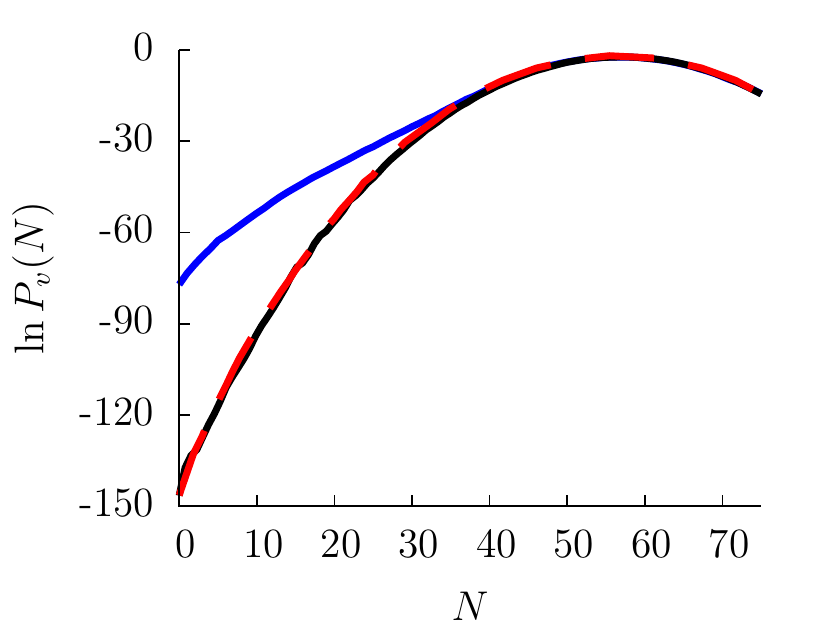}\\
(a)&(b)
\end{tabular}\end{center}
\caption{\label{fig:PvNHydrophobicHydrophilic}Effect of a large solute on the density fluctuations of the surrounding solvent.  (a) The model plate and the probe volume~$\V$.  (b) Probability $\PvN$ of observing $N$~waters in the probe volume~$\V$ in bulk (solid black), next to the hydrophobic plate (solid blue) and next to the hydrophilic plate (dashed red, nearly indistinguishable from solid black). Adapted from Ref.~\cite{PatelVarillyChandler2010}.}
\end{figure}

Liquid-vapor interfaces are soft modes, which is to say that it costs little free energy to deform these structures.  Creating a cavity next to a large solute is thus much easier than creating the same cavity in bulk water.  A demonstration of this effect in water is shown in Fig.~\ref{fig:PvNHydrophobicHydrophilic}.  Here, a model $24\times24\times6\,$\AA\ plate that is either hydrophobic or hydrophilic is placed next to a $24\times24\times3\,$\AA\ probe volume~$\V$, and the probability~$\PvN$ of finding $N$ waters in~$\V$ is calculated.  The hydrophobic plate induces a fat tail in the $\PvN$ distribution, similar to the one depicted in Fig.~\ref{fig:PvNFatTails}.  This occurs because the liquid-vapor interface can be easily deformed to create an empty space inside the probe volume.  Next to the hydrophilic plate, where the interface between water and the plate is not liquid-vapor-like, the $\PvN$ distribution is identical to that obtained when the probe volume $\V$ is far away from the plate.

The fat tails in $\PvN$ distributions are present for large probe volumes in bulk and enhanced by a nearby large solute.  This behavior has two main physical consequences.  First, a hydrophobic object that excludes water from a volume~$\V$ has a lower solvation free energy when next to the large solute than when dispersed in bulk water.  Indeed, the difference in $-\kB T \ln \PvZero$ next to the large solute and in bulk measures the reversible work needed to move the object from a specific position in the bulk to the vicinity of the large solute.  This is precisely the free energy of hydrophobic adhesion.

\begin{figure}
\begin{center}\begin{tabular}{ccc}
\includegraphics{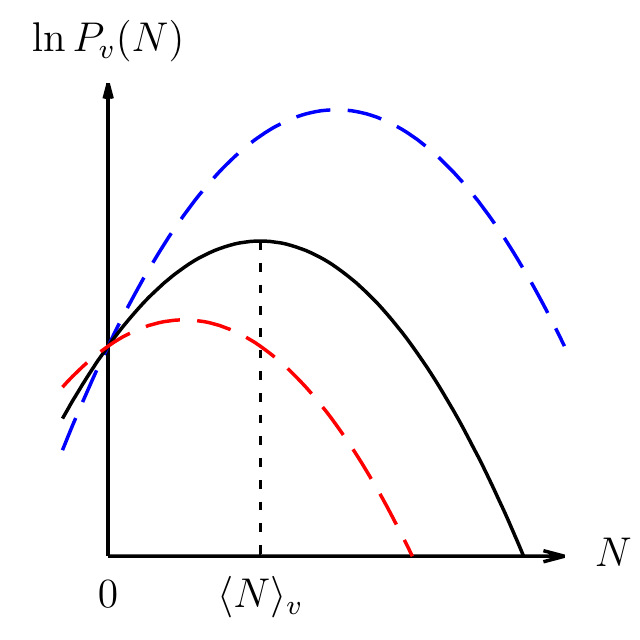}&
\includegraphics{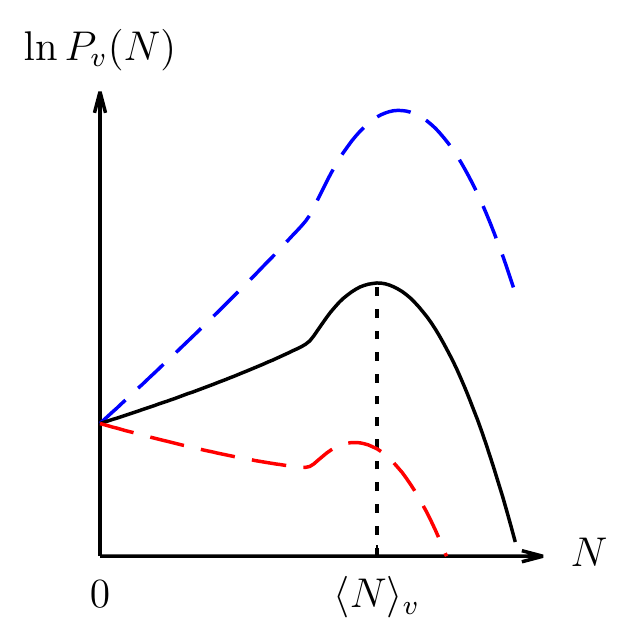}\\
(a) $v$ small&(b) $v$ large
\end{tabular}\end{center}
\caption{\label{fig:FatTailsPerturbed}  Effect of an external perturbation of the form $\epsilon N$ on different $\PvN$ distributions.  (a) When $\V$ is small and $\PvN$ is Gaussian.  The perturbation simply shifts the mean number of waters~$\langle N \rangle_\V$ in~$\V$.  (b) When $\V$ is large, or $\V$ is next to a hydrophobic solute, $\PvN$ has fat tails.  A large enough perturbation results in a first-order microscopic phase transition.  In both figures, the $\PvN$ distribution shown correspond to $\epsilon=0$ (black/solid), $\epsilon < 0$ (blue/dot-dash) and $\epsilon > 0$ (red/dashed).}
\end{figure}

Second, these fat tails also signal an underlying phase instability that can be exposed by an external perturbation.  The idea is illustrated in Fig.~\ref{fig:FatTailsPerturbed}.  Consider an external perturbation of the form~$\epsilon N$.  This perturbation might be the attractive dispersive interactions between a large solute and the water in its vicinity, or the effective repulsion of water from a cavity in confining geometry.  If the $P_\V(N)$ distribution is Gaussian, then a linear perturbation results in another Gaussian free energy of equal width but different mean number of waters in that volume, $\langle N \rangle_\V$.  Small changes in the strength of the perturbation induce small changes in~$\langle N \rangle_\V$.  If, on the other hand, $P_\V(N)$ has a fat tail, then a linear repulsive perturbation can result in a precipitous reduction in~$\langle N \rangle_\V$.  This phenomenon is called a dewetting transition.

This transition can occur in volumes confined by hydrophobic surfaces, such as water inside nano-scale tubes.  Emptying and filling such volumes is thus collective, with several water molecules leaving or entering in bursts.  Refs.~\onlinecite{HummerEtAl2001} and \onlinecite{MaibaumChandler2003} illustrate this behavior.  The phenomenon seems likely relevant to functioning of biological pores.  It is akin to a liquid-vapor transition, but on a scale of nanometers.  Dewetting also appears in cases where two hydrophobic protein surfaces approach one another.  Water confined by these surfaces is destabilized, and as water departs the surfaces move closer together to fill the vacated volume.  This behavior seems relevant to the dynamics of protein folding and assembly, as our group first discussed in Ref.~\onlinecite{tenWoldeChandler2002}.

\subsection{Theory of Dewetting}
Theory for dewetting requires treatment of both interfaces and small length-scale fluctuations.  High free-energetic costs of solvating excluded volumes at small length scales gives way to lower costs in the presence of soft interfacial fluctuations.  Here, we describe how to build a theory that captures this physics with a density field that describes interfaces and a coupling of that field to small-length scale fluctuations.  The development uses some elements of statistical field theory, and while it is therefore a step beyond the simplicity adopted in the earlier parts of our lectures, good textbooks on the topic do exist.  See, for instance, Mehran Kardar's~\onlinecite{Kardar2007b}.

Interfaces are well described by density fields that vary slowly on molecular scales.  In the simplest case, the energetics of such a field, $n(\mathbf{r})$, is given by a Landau-Ginzburg hamiltonian of the form
\begin{equation}
\label{HL}
\beta H_{\mathrm{L}}[n(\mathbf{r})]  = \int \text{d}\mathbf{r} \left[ w(n(\mathbf{r})) + \frac{m}{2} | \nabla n(\mathbf{r}) |^2 \right] \,,
\end{equation}
where we use subscript ``L'' to indicate that this hamiltonian applies to a fluid on \emph{large} length scales only.  The quantity $w(n)$ is a local (grand canonical) free energy density in units of $k_{\mathrm{B}}T = 1/\beta$, and the parameter $m$ determines the free energy cost to create an inhomogeneity.  

In mean field theory, the average $n(\mathbf{r})$ is the function that minimizes this hamiltonian (subject to whatever constraints specify the ensemble considered). This minimization produces a spatially invariant value for $\langle n(\mathbf{r}) \rangle$, except at conditions of phase coexistence where an interface separates volumes with different values of the field.  The interfacial tension for this interface is proportional to $m$ and the shape of the interface is determined by the function $w(n)$.  These relationships can be read about in standard texts, e.g. Ref.~\onlinecite{RowlinsonWidom2003}.

The actual molecular density, $\rho(\mathbf{r})$, can be written as a slowly varying field like $n(\vr)$ plus a correction, where the correction accounts for fluctuations that occur on small-length scales.  In particular, we take
\begin{equation}
\label{Density}
\rho(\mathbf{r}) = n(\mathbf{r}) + \delta \rho(\mathbf{r})\,,
\end{equation}  
where $ \delta \rho(\mathbf{r})$ is the small-length scale part.  There is flexibility in defining this decomposition, but it is important that $n(\mathbf{r})$ varies little over a length $\xi$, which is correlation length of the homogeneous liquid. With this generic criterion, the vapor phase of water is where $n(\mathbf{r})$ is close to zero, and the liquid phase is where $n(\mathbf{r})$ is close to the density of liquid water.  Equation~\eqref{HL} describes the energetics of $n(\mathbf{r})$, and the interface it forms is the liquid-vapor interface.

To the extent that $n(\mathbf{r})$ is a constant equal to the average density of the liquid, $\delta \rho(\mathbf{r}) $ will have the Gaussian hamiltonian, i.e.,
\begin{equation}
\label{HS} 
\beta H_{\mathrm{S}}[\delta \rho(\mathbf{r})] = \frac{1}{2}\int d\mathbf{r} \int d\mathbf{r}' \,\delta \rho(\mathbf{r})\, \chi^{-1}(\mathbf{r},\mathbf{r}')\, \delta \rho(\mathbf{r})
\end{equation}
where $\chi^{-1}(\mathbf{r},\mathbf{r}')$ is the functional inverse of the density-density correlation function, i.e., $\chi(\mathbf{r},\mathbf{r}')= \langle \delta \rho(\mathbf{r})\, \delta \rho(\mathbf{r}')\rangle$.  The subscript ``S'' indicates that this hamiltonian applies to \emph{small}-length scale fluctuations.

The presence of a large enough solute will force the fluid to be inhomogeneous on large length scale.  To account for that possibility, the two fields $n(\mathbf{r})$ and $\delta \rho(\mathbf{r})$ must be coupled, and the simplest way to do so is with a bi-linear form.  Specifically, we take
\begin{equation}
\label{H}
H[n(\mathbf{r}), \delta \rho(\mathbf{r})] = H_{\mathrm{L}}[n(\mathbf{r})] + H_{\mathrm{S}}[\delta \rho(\mathbf{r})]  + H_{\mathrm{I}}[n(\mathbf{r}), \delta \rho(\mathbf{r})]\,, 
\end{equation} 
with
\begin{equation}
\label{HI} 
H_{\mathrm{I}}[n(\mathbf{r}), \delta \rho(\mathbf{r})] = \int d \mathbf{r}\int d \mathbf{r}' \,n(\mathbf{r})\,u(\mathbf{r}, \mathbf{r}')\,\delta \rho(\mathbf{r}') \,+\, H_{\mathrm{norm}}[n(\mathbf{r})],
\end{equation}
where $H_{\mathrm{norm}}[n(\mathbf{r})]$ ensures that 
	\footnote{We leave it as an exercise for the reader to carry out the indicated sum over the Gaussian fields 	to show that  $H_{\mathrm{norm}}[n(\mathbf{r})]$ is an irrelevant constant plus
	
	$$
	\frac{\beta}{2}\int \dee\mathbf{r} \int \dee\mathbf{r}' \int \dee\mathbf{s} \int \dee\mathbf{s}' \,n(\mathbf{r})\,u			(\mathbf{r},\mathbf	{s}) \, \chi(\mathbf{s},\mathbf{s}') \, u(\mathbf{s}',\mathbf{r}')\,n(\mathbf{r}') 	\,. 
	$$}
\begin{equation}
\label{Norm}
 \sum_{\delta \rho(\mathbf{r})} \exp\{-\beta H[n(\mathbf{r}), \delta \rho(\mathbf{r})] \} = \exp\{-\beta H_{\mathrm{L}}[n(\mathbf{r})]\}\,.
 \end{equation}	
The subscript ``I'' labeling $H_{\mathrm{I}}[n(\mathbf{r}), \delta \rho(\mathbf{r})]$ stands for \emph{interaction}, and the symmetric function $u(\mathbf{r}, \mathbf{r}')$ specifies the strength and range of the interaction between the two fields  

While the unperturbed liquid partition sum in Eq.\eqref{Norm} leads back to the Landau-Ginzburg description at large-length scales, a non-trivial alteration occurs in the presence of imposed inhomogeneity.  In the context of possible de-wetting, the most important of these alterations comes from excluded volume $\V$ due to a solute.  This constrains the partition sum in a fashion that is compactly described with the functional
\begin{equation}
\label{Constraint}
C_\V[\rho(\mathbf{r})] =
\begin{cases}
1,&\text{when $\rho(\mathbf{r})=0$ for all $\mathbf{r}\in\V$,}\\
0,&\text{otherwise},
\end{cases}
\end{equation}
where $\V$ denotes the volume that the solute excludes from the solvent.  The volume can be complicated, indeed not even contiguous.  See for instance the excluded volumes depicted in Fig.~\ref{fig:SolvationCartoon} of Lecture~Two.  The excluded volume constraint is common to all solutes, hydrophobic or hydrophilic.  For the latter, a different constraint functional could also be employed, one that binds solvent a regions of space adjoining the excluded volumes. With the constraints imposed by the solute, the partition sum over small-length scale density fluctuations is then
\begin{equation}
\label{WithExclusion}
\sum_{\delta \rho(\mathbf{r})} \exp\{-\beta H[n(\mathbf{r}), \delta \rho(\mathbf{r})] \}\,C_\V[n(\mathbf{r}) + \delta \rho(\mathbf{r})] = \exp\{-\beta \overline{H}_v[n(\mathbf{r})] \}\,,
\end{equation} 
where
\begin{equation}
\label{Hbar}
\overline{H}_v[n(\mathbf{r})] = H_{\mathrm{L}}[n(\mathbf{r})]\,+\,\Delta H_v[n(\mathbf{r})]
\end{equation}\\
The alteration to the Landau-Ginzburg hamiltonian, $\Delta H_v[n(\mathbf{r})]$, is straightforwardly (though tediously) evaluated by carrying out the indicated sum over the Gaussian field in Eq.~\eqref{WithExclusion} \footnote{We leave it as a second exercise to the reader to show that the result of this calculation is
\begin{multline*}
\Delta H_v[n(\mathbf{r})] = -\frac{\kB T}{2} \ln \det (\mathbf{\chi_v^{\mathrm{-1}}}/2\pi)\\
+ \frac{\kB T}{2} \int_v \dee\vr\, \int_v \dee\vr'\,
\biggl[ n(\vr) + \int\dee\vec{s}\, \int\dee\vec{s}'\, \chi(\vr,\vec{s})\, \beta u(\vec{s},\vec{s}')\, n(\vec{s}') \biggr]\\
\chi_v^{-1}(\vr,\vr')%
\biggl[ n(\vr') + \int\dee\vec{s}''\, \int\dee\vec{s}'''\, \chi(\vr',\vec{s}'')\, \beta u(\vec{s}'',\vec{s}''')\, n(\vec{s}''') \biggr],
\end{multline*}
where $\chi_v^{-1}(\vr,\vr')$ is the inverse of $\chi(\vr,\vr')$ when $\vr$~and~$\vr'$ are restricted to the volume~$v$.  In other words, $\chi_v^{-1}(\vr,\vr')$ satisfies
$$
\int_v \dee\vr' \chi_v^{-1}(\vr,\vr') \chi(\vr',\vr'') = \delta( \vr - \vr'' ),\qquad
\text{for $\vr,\vr''\in v$}.
$$
Full details on a method of deriving this result can be found in Ref.~\onlinecite{Chandler1993}.
}.

In using these formulas to compute numbers, some information about the function $u(\mathbf{r}, \mathbf{r}')$ is required.  As it generates a force on the slowly varying field, its functional form need not be very specific.  It suffices to characterize the function in terms of a mean strength and range, i.e.,  $u(\mathbf{r}, \mathbf{r}') = \alpha \, \phi(|\mathbf{r} - \mathbf{r}'|)$, where $\alpha$ is the mean strength with a value of the order of $\kB T$, and $\phi(r)$ is normalized with a range of a few \AA's.  The theory constructed with Eq.\eqref{WithExclusion} is not terribly sensitive to the specific values of strength and range.  Physical arguments can be made to estimate their values \textit{a priori}.  Alternatively, the parameters can be adjusted so that the theory produces essentially perfect agreement between its predictions and the results of a few representative simulation calculations.  A non-zero value of $\alpha$ is needed to capture the wide breadth of the crossover from small- to large-length scale behaviors, e.g., as depicted in Fig.~\ref{fig:SolvationScaling}.

A subtlety in these formulas concerns the form of $\chi(\mathbf{r}, \mathbf{r}')$.  When $n(\mathbf{r})$ is not a constant, this variance is not simply the liquid phase function given in Eq.~\eqref{PairCorrelation} of Lecture~Two.  A simple interpolation formula can be used to estimate the variance of the small-length scale field,
\begin{equation}
\chi(\mathbf{r}, \mathbf{r}') \approx  n(\mathbf{r})\,\delta(\mathbf{r} - \mathbf{r}')\,+\,n(\mathbf{r}\,)\,[g(|\mathbf{r}-\mathbf{r}'|)-1]\,n(\mathbf{r}').
\end{equation}
Notice that this formula guarantees that there are no density fluctuations wherever $n(\mathbf{r})=0$.  Again, because $n(\mathbf{r})$ is slowly varying, more molecular-scale detail than provided reliably by this interpolation formula is unimportant to the evaluation of $\Delta H_v[n(\mathbf{r})]$.

This evaluation establishes the following behaviors:  
\begin{enumerate}
\item If the excluded volume $\V$ is not much larger than the correlation volume of the unperturbed liquid, $\xi^3$, or if it is composed of several distantly separated excluded volumes, each one not much larger  than a correlation volume, then $\Delta H_v[n(\mathbf{r})]$ is essentially a constant and equal to the solvation free energy of the excluded volume as given by Gaussian fluctuation theory, Eq.~\eqref{eqn:DeltaMuSmall} in Lecture~Two. 

\item On the other hand, if $\V$ presents a surface of low curvature extending over lengths larger than $\xi$, then $\Delta H_v[n(\mathbf{r})]$ is no longer constant and is relatively large when $n(\mathbf{r}) \neq 0$ for $\mathbf{r} \in \V$.  To avoid this energetic penalty, probable configurations of the slowly varying fields will adjust to make $n(\mathbf{r}) = 0$ within the excluded volume.  As such, according to Eq.~\eqref{HL}, the probability for configurations of the slowly varying field will be maximal at configurations with a liquid-vapor-like interface adjacent to the excluded volume. 

\end{enumerate}

This latter situation is that of dewetting.  Regions of space where dewetting occurs is where an excluded volume $\V$ causes the average value of $n(\mathbf{r})$ to be zero.  It occurs only for cases of sufficiently large excluded volumes.  The field $n(\mathbf{r})$ governed by Eq.~\eqref{HL} is a continuum version of an Ising model or lattice-gas model.  We see from the theory sketched above that the underlying physics of dewetting is captured by coupling of an Ising-like field to a small-length scale field through excluded volume perturbations.  Indeed, with a single fixed parameter the strength parameter $\alpha$, the results of these equations agree quantitatively with those of computer simulations, as we have recently shown in detail in Ref.~\onlinecite{VarillyPatelChandler2011}.  

\subsection{Applications and hydrophobic collapse}  
The first general theoretical treatment of dewetting in water and its role in hydrophobicity was provided by the Lum-Chandler-Weeks (LCW) theory~\cite{LumChandlerWeeks1999}.  That theory is the mean-field approximation to what we have presented above.  In particular, 
\begin{equation}
\label{LCW}
\langle n(\mathbf{r}) \rangle \approx n_{\mathrm{LCW}}(\mathbf{r})\,,
\end{equation} 
where $ n_{\mathrm{LCW}}(\mathbf{r})$ is the field that minimizes $\overline{H}_v[n(\mathbf{r})] $, i.e.,
\begin{equation}
\label{LCWmin}
0 = \delta \overline{H}_v[n_{\mathrm{LCW}}(\mathbf{r})] / \delta n_{\mathrm{LCW}}(\mathbf{r})\,.
\end{equation}
The LCW paper~\cite{LumChandlerWeeks1999} presents several illustrative predictions of Eqs.~\eqref{LCW} and~\eqref{LCWmin}.  These predictions have motivated many subsequent studies of hydrophobic effects.

More recent work building from this approach have focused on fluctuations in $n(\mathbf{r})$.  These fluctuations are important in dynamics.  The weight functional is
\begin{equation}
\label{PartitionFunction}
P[n(\mathbf{r})] \propto \exp\{-\beta \overline{H}_v[n(\mathbf{r})]\}\,,
\end{equation}
which can be sampled by Monte Carlo.  The simplest implementation replaces $n(\mathbf{r})$ with a binary field on a lattice, $\rho \,n_i$, where $\rho$ is the bulk liquid density and $n_i$ is either 0 or 1 depending upon whether $i$th cell in the lattice is vapor-like or liquid-like, respectively.  $H_{\mathrm{L}}[n(\mathbf{r})]$ is then taken to be a lattice-gas hamiltonian, and $\Delta H_v[n(\mathbf{r})] \approx \sum_i c\, n_i v_i$, where $c$ is a positive constant, and $v_i$ is the volume excluded by solutes in cell $i$.  That is, 
\begin{equation}
\label{Simplest}
\overline{H}[n(\vr)] \approx - \epsilon \sum_{i,j}{\,}^{'} \,n_i\,n_j\, +\,\sum_i (c\, v_i - \rho\, \mu)\,n_i\,,
\end{equation}
where $\mu$ is the chemical potential of the liquid and the primed sum is over nearest neighbors.  The $c\,v_i$ terms account for excluded volume in liquid-like regions, with the free energy cost for this excluded to be proportional to the size of that volume.  Proportionality to excluded volume is approximately consistent with small-length scale hydrophobic free energies of solvation, as we explained in Lecture~Two.

This version of the theory is particularly easy to implement.  With judicious choices of lattice spacing and the constant $c$, most qualitative features of dewetting and hydrophobic forces of assembly are captured correctly.  The principal feature it fails to capture is the slow approach to the macroscopic surface-area scaling illustrated in Fig.~\ref{fig:SolvationScaling}.  Instead of a broad crossover from small to large-length scale hydrophobicity, this simplest version exhibits a relatively abrupt crossover around 1 nm.  This deficiency can be ameliorated, as we have detailed in Ref.~\onlinecite{VarillyPatelChandler2011}, but the simplicity of the model described with Eq.~\eqref{Simplest} makes it an attractive choice for easy estimates of roles of hydrophobic forces.

\begin{figure}
\begin{center}\includegraphics[width=130mm]{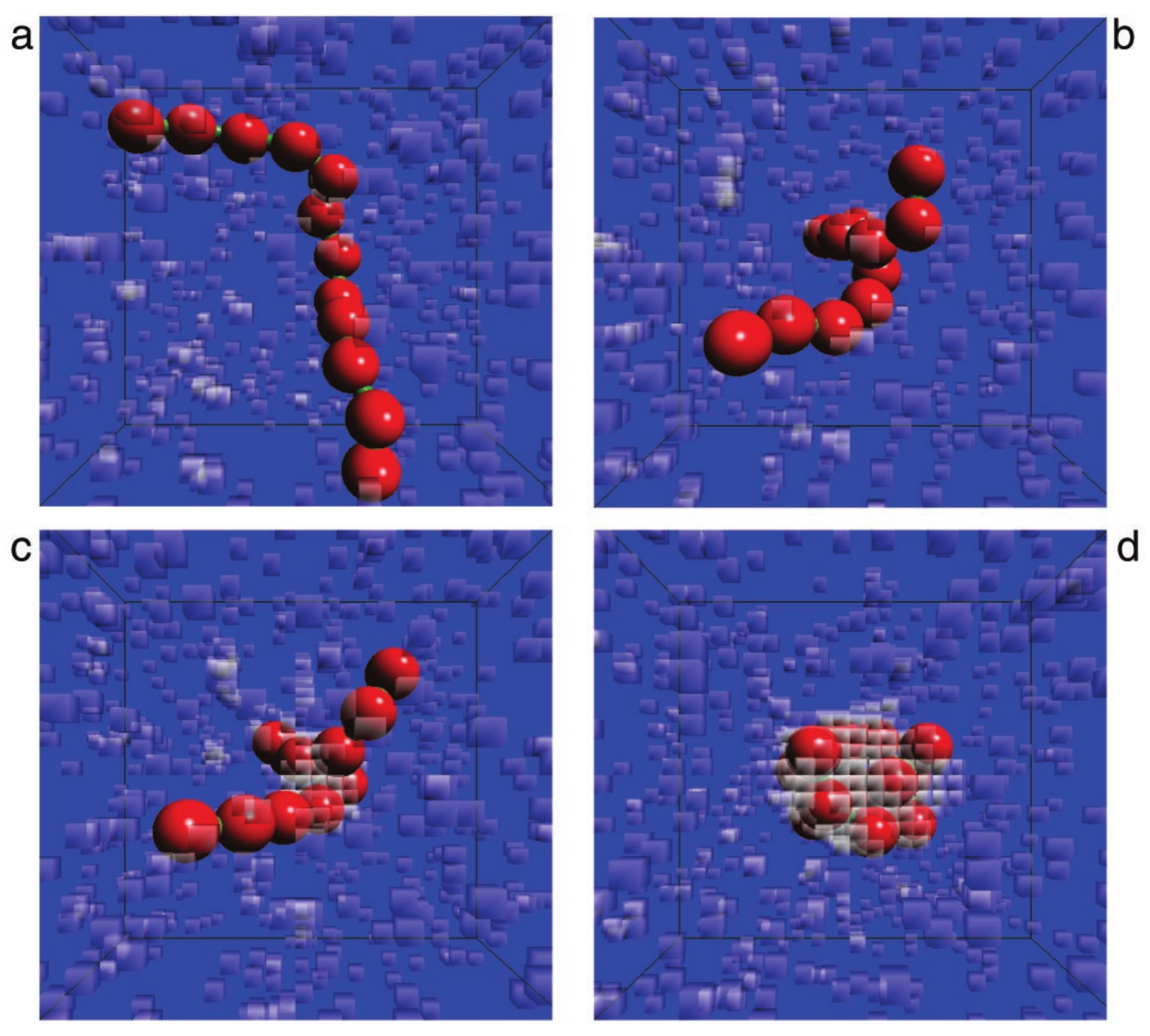}\end{center}
\caption{\label{fig:tWC02Trajectory}Example collapse trajectory of a model hydrophobic polymer embedded in solvent.  Frame~(a) shows a typical extended configuration.  Frames (b), (c)~and~(d) are snapshots from a~$1.5\,$ns collapse trajectory, with the configuration shown in Frame~(c) being a transition state.   The white cubes are the cells in the lattice gas where $n_i$ is~$0$.  In this model, the cell size $\ell$ is chosen to be~$2.1\,$\AA.  From Ref.~\onlinecite{tenWoldeChandler2002}.}
\end{figure}

Figure~\ref{fig:tWC02Trajectory} shows a result obtained from this form of modeling. Depicted are snap shots of a trajectory illustrating the collapse of a chain of hard spheres in water.  The chain follows Newtonian dynamics with friction and random forces reflecting the effects of the small-length scale field that has been integrated out.  The liquid's dynamics is Monte Carlo.  Motion of the chain is reflected in changes of $v_i$, and as such the liquid's slowly varying field  couples to the dynamics of the chain.  The two move together illustrating the collective nature of hydrophobic forces of assembly. 

The intra-chain forces for the chain considered in that figure are such that the extended chain is most stable configuration in the gas phase. Solvation, in this case hydrophobic forces of assembly, make the globule state the most stable configurations in the liquid.  The half-life of the extended chain in the liquid is about 1~$\mu$s.  The pictured trajectory shows parts of a trajectory during the relatively short period of time when the chain passes from its metastable extended state to the stable globule state.  

The transition state occurs when a random fluctuation of the chain produces a large enough cluster of hard spheres to nucleate a soft liquid-vapor-like interface.  At that stage, water moves away with relatively low free energy cost and the chain collapses to its most stable thermodynamic state.  In effect, the reorganization of the chain boils away the water.  It is a suggestive of the possibility of a nano-scale steam engine, albeit in a much more complicated device than a simple hydrophobic chain.  It is a possibility that we ponder and wonder if it exists in some biological molecular motor.  
 
\end{document}